\documentclass[aps,prc,10pt,eqsecnum,twocolumn,dvips,showpacs]{revtex4-1}
\usepackage{graphicx}
\usepackage[tbtags]{amsmath}
\usepackage{bm,amssymb}
\usepackage{mathptm}
\def\br{{\bm r}}
\def\bk{{\bm k}}

\def\<{\langle}
\def\>{\rangle}

\newcommand{\pp}[2]{\frac{\partial #1}{\partial #2}}
\renewcommand{\Re}{\mathop{\rm Re}}
\renewcommand{\Im}{\mathop{\rm Im}}
\newcommand{\NN}{\nonumber\\}
\newcommand{\etal}{{\it et al.}}

\begin{document}
\title{Semiclassical trace formula for 
truncated spherical well potentials: \\
Toward the analyses of shell structures in nuclear fission processes}

\author{Ken-ichiro Arita}
\affiliation{Department of Physics,
Nagoya Institute of Technology, Nagoya 466-8555, Japan}
\begin{abstract}
Trace formulas for the contributions of degenerate periodic-orbit
families to the semiclassical level density in truncated spherical
hard-wall potentials are derived.  In addition to the portion of the
continuous periodic-orbit family contribution which persists after
truncation, end-point corrections to the truncated family should be
taken into account.  I propose a formula to evaluate these end-point
corrections as separate contributions of what I call marginal orbits.
Applications to the two-dimensional billiard and three-dimensional
cavity systems with the three-quadratic-surfaces shape
parametrization, initiated to describe the nuclear fission processes,
reveal unexpectedly large effects of the marginal orbits.
\end{abstract}
\maketitle

\section{Introduction}
In quantum many-body systems such as nuclei and microclusters, the
fluctuations in the physical quantities like energy and deformations
as functions of the constituent particle number are essentially
governed by the single-particle shell effects.  In many cases, gross
structures in the single-particle energy spectra show regularly
oscillating patterns.  The origins of such patterns are clearly
explained using the semiclassical periodic-orbit theory (POT), which
expresses the quantum level density and also shell energy in terms of
the contributions of classical periodic orbits.  This formula, known
as the trace formula, immediately after its discovery by
Gutzwiller\cite{Gutz71,GutzText} and independently by
Balian-Bloch\cite{BB1,BB3} was recognized as being extremely useful in
explaining the properties of gross shell structures in nuclei, shell
and supershell structures in metallic clusters, and so on.  Bunchings
of levels in isotropic harmonic oscillator system or in anisotropic
ones with rational frequency ratios are related to the conditions for
most of the classical orbits to be periodic with short periods, which
qualitatively explain the origins of spherical and superdeformed shell
structures\cite{BM2}.  Strutinsky and coworkers applied the trace
formula to explain the deformed shell structures in the nuclear mean
field\cite{StrMag76,StrMag77}.  Nishioka \etal have set out an elegant
explanation of the supershell structures in metallic clusters as the
interference of the contributions of triangle and square type periodic
orbits in the spherical mean field potential\cite{Nishioka}.

In the low-energy fission of actinide nuclei, the double-humped
structure in the fission barrier and the asymmetric fragment-mass
distributions are caused by the quantum shell effect\cite{FunnyHill}.
According to the POT, shell structures associated with the periodic
orbits which oscillate twice along the minor axis while they oscillate
once along the major axis in a strongly elongated mean-field potential
play a significant role in building the double-humped fission
barrier\cite{Arita98,Magner2002,Magner2011}.  Brack \etal have made a
semiclassical analysis of the origin of the asymmetric fission using a
simple cavity potential model\cite{Brack97}.  They have found that the
valleys in the potential energy surface from symmetric minima toward
strongly elongated asymmetric shapes can be nicely explained by the
contribution of the shortest periodic orbit.

More specifically, the fragment-mass distribution in low-energy
nuclear fission experiments suggests strong effects of the fragment
shell structures.  In the fission of actinide nuclei, the mass numbers
of the heavier fragments are around 140, independent of the mass of
the parent nuclei, which is close to that of the doubly-magic
$\mathrm{^{132}Sn}$ $(Z=50, N=82)$.  On the other hand, the recent
experiment on the fission of neutron-deficient mercury isotope
$\mathrm{^{180}Hg}$ shows asymmetric fragment-mass distribution in
spite of the stability of the fragments $\mathrm{^{90}Zr}$ $(Z=40,
N=50)$ in the case of symmetric fission\cite{Andr2012}.  According to
the theoretical analysis of the five-dimensional potential energy
surface with the macroscopic-microscopic model\cite{Ichikawa2012}, the
hindrance of symmetric fission in $\mathrm{^{180}Hg}$ is successfully
interpreted as the result of a large potential barrier along the
symmetric path in the potential energy surface, which stems mainly
from the shell effect.  This implies the importance of the deformed
shell effect in the fission process, rather than those of the final
daughter nuclei, to understand the properties of the fragment-mass
distributions.

For the nucleus in the fission process, a neck is formed in the
mean-field potential and it gradually separates the system into two
nascent fragment parts.  In the following, I use the term
``prefragment'' denoting the nascent fragment to distinguish it from
those after the scission.  When the neck is developed, one may expect
a kind of shell effect which stabilize the shapes and sizes of the
prefragment parts.  An unexpectedly large prefragment shell effect at
a rather early stage of the fission process, just after getting over
the second saddle, was suggested by the two-center shell model
calculation\cite{Mosel_NPA71}.  Emergence of the prefragments, which
have density profiles similar to those of stable spherical magic
nuclei, was found in modern microscopic density-functional
calculations\cite{Warda2012,McDonnell2013}.  Although those results
suggest the significance of shell effects associated with the
prefragments, it is not a simple problem to extract the effect of each
prefragment exclusively out of the total shell effect in purely
quantum mechanical approaches, because most of the single-particle
wave functions are delocalized in the potential.

Here, let us take notice of the fact that the semiclassical level
density is represented by sum of the contributions of classical
periodic orbits.  Formation of neck (constriction) in the potential
yields periodic orbits which are confined in each of the prefragment
parts, which will be briefly termed as ``prefragment orbits'' below.
Then, one can define the prefragment shell effects unambiguously by
the contributions of those prefragment orbits to the semiclassical
level density.  Note that those orbits bring about the same kinds of
quantum fluctuation to the system as in the case where they exist in
an isolated fragment.  Taking account of those features, one can
analyze the roles of the neck formation in stabilizing the shapes of
nuclei in the fission processes.

In order to focus on the effect of shape evolution, a simple cavity
potential model will be employed but with the ingenious
three-quadratic-surfaces (TQS) shape parametrization, which is useful
in describing the nuclear fission processes.  In this parametrization,
two prefragments and the neck part between them are represented by
quadratic surfaces, and their shapes are easily controlled by the
shape parameters.  Supposing that the prefragment has a spherical
shape, which provides the strongest shell effect, one must necessarily
treat the classical periodic-orbit families confined in the truncated
spheres for the semiclassical analysis of the deformed shell effect.
Periodic orbits which form a continuous family having identical action
and stability are called degenerate, and the order of degeneracy is
defined by the number of independent continuous parameters for the
family.  The polygon and diameter orbits form three- and two-parameter
families in the truncated spherical cavity, but the ranges of the
parameters are restricted and their contributions to the semiclassical
level density are suppressed compared with those in the nontruncated
spherical cavity.

In this paper, I derive the contribution of degenerate periodic-orbit
family confined in the truncated spherical cavity potential based on
the Balian-Bloch formula\cite{BB3}.  For simplicity, I begin with a
two-dimensional (2D) billiard system using the same shape
parametrization.  In Sec.~\ref{sec:theory}, the essence of the
Balian-Bloch trace formula is briefly outlined, and then contributions
of degenerate families of periodic orbits in 2D truncated circular
billiard and a three-dimensional (3D) truncated spherical cavity are
derived.  As numerical applications, I examine the single-particle
level densities in the TQS billiard and cavity models in
Sec.~\ref{sec:applications}.  Section~\ref{sec:summary} is devoted to
a summary and concluding remarks.  Some details in derivations of the
trace formulas are given in the Appendix.

\section{Trace formula for degenerate orbits in hard-wall potentials}
\label{sec:theory}

\subsection{Balian-Bloch formula}

I shall first outline the derivation of the Balian-Bloch formulas for
semiclassical level density in hard-wall potential
models\cite{BB1,BB3}.  Consider a particle of mass $M$ which moves
freely inside the closed surface $S$ and is reflected ideally on the
wall.  The energy of the particle is given by $E=\hbar^2k^2/2M$ with
the constant wave number $k$.  The Green's function for such a system
is defined by
\begin{gather}
\left(-\frac{\hbar^2}{2M}\bm{\nabla}^2-E\right)
G\left(\br,\br';E\right)=\delta(\br-\br'),
\end{gather}
with the Dirichlet boundary condition $G(\br_s,\br';E)=0$ for $\br_s$
on the wall $S$ and $\br'$ inside the wall.  In terms of the Green's
function, level density $g(E)$ is expressed as
\begin{equation}
g(E)=\frac{1}{\pi}\Im\int_V d\br\; G(\br,\br;E+i0)
\label{eq:ld_green}
\end{equation}
where the volume integral is taken over the interior region $V$ of the
closed surface $S$.  By introducing a double-layer potential on the
surface to ensure the boundary condition\cite{CHVol2}, a
multiple-reflection expansion formula for the Green's function is
derived\cite{BB1}, which is expressed as
\begin{gather}
G(\br_0,\br_0';E)=G_0(0,0')+\sum_{p=1}^\infty\left(\frac{\hbar^2}{M}\right)^p
\oint_S dS_1\cdots dS_p \NN \times \pp{G_0(0,p)}{n_p}
\pp{G_0(p,p-1)}{n_{p-1}}\cdots\pp{G_0(2,1)}{n_1}G_0(1,0').
\label{eq:multi_reflection}
\end{gather}
Here, $G_0(b,a)=G_0(\br_b,\br_a;E+i0)$ denotes the Green's function
for a free particle, and $\partial/\partial n_a$ represents the
component of the gradient normal to the surface $S$ at $\br_a$.  Each
term on the right-hand side can be interpreted as the contribution of
the wave which starts off at $\br_0'$ and hits $p$ times on the wall
$S$ at $\br_1,\cdots, \br_p$ before arriving at $\br_0$.  Substituting
(\ref{eq:multi_reflection}) into (\ref{eq:ld_green}), one has
\begin{gather}
g(E)=g_0(E)+\frac{1}{\pi}\Im\sum_{p=1}^\infty
\left(\frac{\hbar^2}{M}\right)^p\int_V d\br_0
\oint_S dS_1\cdots dS_p \NN \quad \times \pp{G_0(0,p)}{n_p}
\pp{G_0(p,p-1)}{n_{p-1}}\cdots\pp{G_0(2,1)}{n_1}G_0(1,0)
\end{gather}
For sufficiently large $k$, integrations on the right-hand side can be
carried out using the stationary-phase approximation (SPA), and the
level density is expressed as the sum over contributions of the
stationary paths, namely, the classical periodic orbits.  The
free-particle Green's function $G_0$ is given by
\begin{align}
&\frac{\hbar^2}{2M}G_0(\br_b,\br_a;E) \NN
&=\left\{
\begin{array}{l@{\quad}l}
\dfrac{i}{4}H_0^{(1)}(kr_{ab})\simeq
 \dfrac{e^{ikr_{ab}}}{\sqrt{8\pi kr_{ab}/i}}, & \text{(2D)} \\
\dfrac{e^{ikr_{ab}}}{4\pi r_{ab}}, & \text{(3D)}
\end{array}\right.
\label{eq:green_free}
\end{align}
for the spatial dimensions 2 and 3, with $r_{ab}=|\br_b-\br_a|$.  In
the expression for 2D billiard, $H_\nu^{(1)}$ denotes the $\nu$th
order Hankel function of the first kind, and the approximation on the
right-hand side holds for the asymptotic limit $kr_{ab}\gg 1$.  Using
these expressions, one arrives at the general formula for the
semiclassical level density
\begin{align}
g(E)=g_0(E)+ & \Re\sum_\beta a_{\beta}(k)
\oint dS_1\cdots dS_{p_\beta} e^{ikl_\beta}.
\label{eq:trace_gen}
\end{align}
$l_\beta$ denotes the total length of the polygon orbit with
$p=p_\beta(\geq 2)$ vertices on the wall $S$,
\begin{equation}
l_p=r_{12}+r_{23}+\cdots+r_{p-1,p}+r_{p1},
\end{equation}
which is expressed as a function of the local surface coordinates
around the vertices of the stationary orbit $\beta$.  The
preexponential factor is evaluated for the stationary orbit $\beta$
and is put out of the integral into $a_\beta(k)$ as usual in the SPA
(see Ref.~\cite{BB3} for its explicit form in the 3D case).  For a
system with only isolated periodic orbits, all the surface integrals
in (\ref{eq:trace_gen}) are carried out by expanding the length $l_p$
with respect to the surface coordinates up to the second order around
the stationary point, and the integrals are reduced to the Fresnel
type.  The result can be translated to the Gutzwiller trace formula
\cite{Gutz71,GutzText}
\begin{gather}
g(E)=g_0(E)+\sum_\beta\sum_{m=1}^\infty
\frac{T_\beta}{\pi\hbar\sqrt{|\det(\mathsf{M}_\beta^m-I)|}} \NN
\times\cos\Bigl(mkL_\beta-\tfrac{\pi}{2}\mu_{m\beta}
\Bigr). \label{eq:Gutz_trace_E}
\end{gather}
On the right-hand side, the sum is taken over all the primitive
periodic orbits $\beta$ and the numbers of their repetitions $m$
($m=1$ corresponds to the primitive orbit).  $T_\beta$ is the period
of the primitive orbit $\beta$,
\begin{equation}
T_\beta=\frac{dS_\beta}{dE}=\frac{ML_\beta}{\hbar k},
\end{equation}
with the wave number $k=\sqrt{2ME}/\hbar$ and the orbit length
$L_\beta$.  In Eq.~(\ref{eq:Gutz_trace_E}), $\mathsf{M}_\beta$
represents the monodromy matrix which describes the stability of the
orbit, and $\mu_{m\beta}$ is the Maslov index related to the number of
focal and caustic points along the orbit\cite{Gutz71,GutzText}.  The
level density in terms of the wave-number variable $k$ is written as
\begin{align}
g(k)&=g(E)\frac{dE}{dk} \NN
&=g_0(k)+\sum_\beta\sum_{m=1}^\infty
\frac{L_\beta}{\pi\sqrt{|\det(\mathsf{M}_\beta^m-I)|}} \NN
&\qquad\times\cos\Bigl(mkL_\beta-\tfrac{\pi}{2}\mu_{m\beta}\Bigr).
\label{eq:Gutz_trace}
\end{align}

In a Hamiltonian system with continuous symmetries, generic periodic
orbits form continuous families generated by the symmetry
transformations.  For such degenerate periodic orbits, some of the
integrals in (\ref{eq:trace_gen}) should be carried out exactly with
respect to the continuous parameters for the family.  The extensions
of the Gutzwiller trace formula to systems with continuous symmetries
are presented in \cite{StrMag76,Creagh91,Creagh92}.

In a 2D circular billiard with radius $R_0$, there are regular polygon
orbits labeled by the two integers $(p,t)$, where $p$ is the number of
vertices and $t$ is the number of turns around the center $(p\geq
2t)$.  A primitive orbit is specified by an incommensurable pair of
$p$ and $t$, and the repeated orbit with repetition number $m$ is
denoted by $m(p,t)$.  Each of those orbits forms a one-parameter
family due to the rotational symmetry.  Then, integrals in
Eq.~(\ref{eq:trace_gen}) are done for the one surface coordinate
associated with the degeneracy exactly, and for the others using the
SPA.  The analytic expression of the result for the orbit $(p,t)$ is
obtained as
\cite{Reimann96,BrackText}
\begin{gather}
g_{m(p,t)}^{\rm(circ)}(k)=2R_0\sqrt{kR_0}A_{m(p,t)}^{\rm(circ)}
\sin\left(kmL_{pt}-\tfrac{\pi}{2}\mu_{m(p,t)}^{\rm(circ)}\right),
\label{eq:trace_circ}
\end{gather}
with the dimensionless energy-independent amplitude factor $A$, orbit
length $L$, and the Maslov index $\mu$ given by
\begin{gather}
A_{m(p,t)}^{\rm(circ)}=w_{pt}
\sqrt{\frac{\sin^3\varphi_{pt}}{mp\pi}}, \quad
\varphi_{pt}=\frac{\pi t}{p}, \NN
L_{pt}=2pR_0\sin\varphi_{pt}, \quad
\mu_{m(p,t)}^{\rm(circ)}=3mp-\tfrac32.
\end{gather}
Here, $w_{pt}$ represents the time-reversal factor: It takes the value
$2$ for polygon orbits $(p>2t)$ to take into account the orbits
turning clockwise and anticlockwise, while it takes the value $1$ for
diameter orbits $(p=2t)$ whose time reversals are equivalent to the
original ones.

In a 3D spherical cavity potential with radius $R_0$, there exist the
same set of periodic orbits as in the circular billiard but with
different degeneracies\cite{BB3}.  Polygon orbits $(p>2t)$ form
three-parameter families generated by the three-dimensional rotations.
To obtain the contribution of such a family, the integrals in
Eq.~(\ref{eq:trace_gen}) are done for three surface coordinates
associated with the degeneracy exactly, and for others by using the
SPA.  The analytic expression is obtained as\cite{BB3}
\begin{gather}
g_{m(p,t)}^{\rm(sph)}(k)=2R_0(kR_0)^{3/2}A_{m(p,t)}^{\rm(sph)}
\sin\Bigl(kmL_{pt}-\tfrac{\pi}{2}\mu_{m(p,t)}^{\rm(sph)}\Bigr),
\label{eq:trace_sphere}
\end{gather}
with
\begin{gather}
A_{m(p,t)}^{\rm(sph)}=\sin(2\varphi_{pt})
\sqrt{\frac{\sin\varphi_{pt}}{mp\pi}},\quad
\mu_{m(p,t)}^{\rm(sph)}=m(2t-p)-\tfrac32. \label{eq:BB_sph}
\end{gather}
On the other hand, the diameter orbit $(p=2t)$ forms a two-parameter
family since the rotation about the diameter itself does not generate
a family.  The contribution of the diameter family $m(2,1)$ is also
derived from Eq.~(\ref{eq:trace_gen}) in the same manner as the
polygon orbits, and is expressed as\cite{BB3}
\begin{gather}
g_{m(2,1)}^{\rm(sph)}(k)
=2R_0(kR_0)A_{m(2,1)}^{\rm(sph)}
\sin\Bigl(kmL_{21}-\tfrac{\pi}{2}\mu_{m(2,1)}\Bigr),
\label{eq:trace_sph2}
\end{gather}
with
\begin{gather}
A_{m(2,1)}^{\rm(sph)}=\frac{1}{2\pi m}, \quad \mu_{m(2,1)}=2.
\end{gather}

\begin{figure}[tb]
\centering
\includegraphics[width=\linewidth]{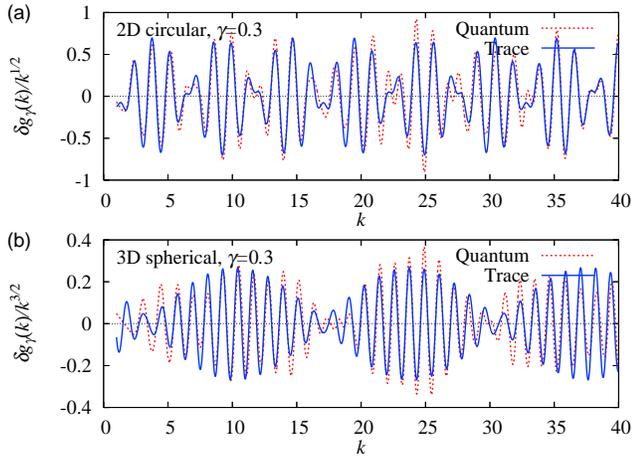}
\caption{\label{fig:sld0} 
Oscillating part of the level density with averaging width $\gamma=0.3$
for (a) 2D circular billiard and (b) 3D spherical cavity systems.  In
each panel, the dotted (red) line represents the quantum result and
the solid (blue) line represents the result of the semiclassical trace
formula with the contributions of a few major orbits.  For the
circular billiard, the contributions of the diameter (2,1) and
triangle (3,1) orbit families are taken into account.  For the
spherical cavity, the contributions of the triangle (3,1) and square
(4,1) orbit families are considered.}
\end{figure}

In general, gross shell structure is governed by the contribution of
some shortest periodic orbits.  In Fig.~\ref{fig:sld0}, the
oscillating part of the level density averaged with the width
$\gamma$,
\begin{equation}
\delta g_\gamma(k)=\int dk'[g(k')-\bar{g}(k')]
\exp\left[-\frac12\left(\frac{k-k'}{\gamma}\right)^2\right],
\label{eq:sld_gam}
\end{equation}
is shown for the 2D circular billiard and the 3D spherical cavity.
The smoothing width $\gamma=0.3$ is taken, for which only the orbits
with length $L\lesssim \pi/\gamma\approx 10$ contribute.  In the 2D
circular billiard, all the periodic orbits form a one-parameter family
and the dominance of their contribution to the gross shell structure
is mainly determined by the shortness of the length.  In the upper
panel of Fig.~\ref{fig:sld0}, one sees that the quantum result of
$\delta g_\gamma(k)$ for 2D circular billiard is nicely reproduced by
the semiclassical formula (\ref{eq:trace_circ}) with the contribution
of only two shortest orbits, diameter (2,1) and triangle (3,1).  In
the 3D spherical cavity, the shortest orbit is the diameter, but it
plays a minor role compared with the other polygon families due to the
low degeneracy.  In the lower panel of Fig.~\ref{fig:sld0}, the
quantum result of $g_\gamma(k)$ for the 3D spherical cavity is
compared with the semiclassical trace formula (\ref{eq:trace_sphere})
taking the contributions of only triangle (3,1) and square (4,1)
families into account.  One sees that the outstanding beating pattern
called supershell structure is successfully reproduced as the
interference effect of those two orbits\cite{BB3}.  The agreements of
the semiclassical trace formula with the quantum results are already
fine with the above two main orbits, and become much better when the
contributions of other remaining orbits are incorporated.

\subsection{Two-dimensional truncated circular billiard}

Now I consider a 2D billiard system with the wall partly consisting of
a circular arc whose central angle is larger than $\pi$.  In such a
billiard potential, one has a degenerate family of diameter orbits
confined in the circle part.  In general, there exists a family of
regular polygon orbits with $p$ vertices when the central angle of the
arc is larger than $2\pi(1-\frac{1}{p})$.
\begin{figure}[tb]
\includegraphics[width=.6\linewidth]{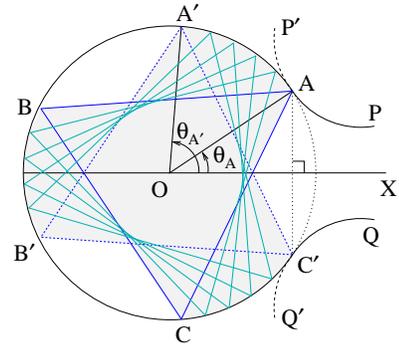}
\caption{\label{fig:opencirc}
Degenerate family of triangle orbits in the truncated circular
billiard.  The circular wall is truncated at the points A and
$\mathrm{C'}$, where the wall is smoothly connected to the outer parts
AP and $\mathrm{C'Q}$.  A continuous family of orbits $(p,t)$ is
available for the angles $\theta_A\equiv\angle\mathrm{AOX} <\pi/p$,
where OX is the axis of symmetry for the circle wall.  For instance,
the equilateral triangle orbit (3,1) is possible for $\theta_A<\pi/3$
and forms a continuous family by rotating anticlockwise from ABC to
$\mathrm{A'B'C'}$.}
\end{figure}
As an example, shown in Fig.~\ref{fig:opencirc}, the rotation around
the center O generates a continuous family of triangle orbits (3,1)
ranges from ABC to $\mathrm{A'B'C'}$.  To avoid the complication due
to singularities, I assume that the circle part of the wall is
smoothly connected to the neighboring walls AP and $\mathrm{C'Q}$ as
illustrated in Fig.~\ref{fig:opencirc}.

\begin{figure}[t]
\centering
\includegraphics[width=\linewidth]{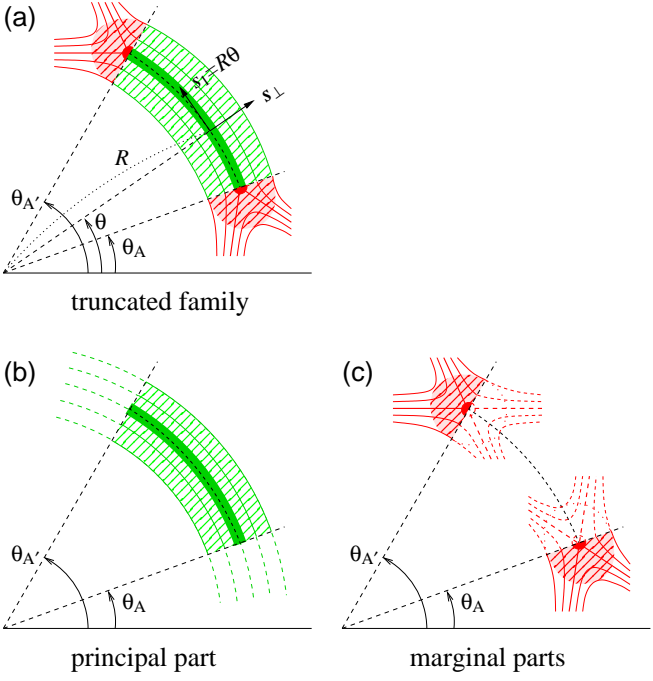}
\caption{\label{fig:marginal}
Illustration of the distribution of periodic orbits on the surface
coordinate space for truncated circular billiard [panel (a)].  Thin
solid lines represent the contour of the orbit length
$l_{pt}(\bm{s})$, the thick line represents the degenerate
periodic-orbit family, and solid dots at the both ends of the family
denote the marginal orbits.  In the stationary-phase approximation,
the shaded area around the family (set of nearly periodic orbits)
makes a major contribution to the integral (\ref{eq:trace_gen}).  In
the above area, the part $\theta_{\rm A}<\theta<\theta_{\rm A'}$
[panel (b)] gives the principal term (\ref{eq:trace_pr}), and the
terminal semi-disc parts around the dots [panel (c)] give the marginal
orbit contributions (\ref{eq:trace_mg}).  The broken lines in panel
(c) represent the contour of the orbit length in the case when the
wall outside of the circle part is extended toward the inside as AP',
illustrated in Fig.~\ref{fig:opencirc}.}
\end{figure}

Let us consider the contribution of this orbit family to the
semiclassical level density based on Eq.~(\ref{eq:trace_gen}).  Fixing
the position of the first vertex $\mathrm{P_1}$, the positions of the
other vertices of the stationary path are uniquely determined.  I take
$s_1=R\theta$ as the position of $\mathrm{P_1}$ and $s_j$ $(j\geq 2)$
as the displacement of the $j$th vertex from its position in the
stationary path for given $s_1$.  Figure~\ref{fig:marginal}
schematically shows the distribution of periodic orbits on the surface
coordinate space $\bm{s}=(s_1,\cdots, s_p)$.  Panel (a) shows a
schematic contour plot of the length $l_{pt}$ near the truncated
periodic-orbit family.  The stationary points of the length $l_{pt}$
give the periodic orbits, and the thick solid line represents the
continuous set of stationary points corresponding to the degenerate
periodic-orbit family parametrized by the rotation angle $\theta$.  It
is truncated at $\theta=\theta_{\rm A}$ and $\theta_{\rm A'}$
indicated by the dots, which correspond to what I call the marginal
orbits.  I evaluate the integrals in Eq.~(\ref{eq:trace_gen}) by
dividing the integration range into two parts: the interior portion
(principal term) and the area around the end points (marginal term).
In the interior portion of the family, $\theta_{\rm A}<\theta
<\theta_{\rm A'}$, the orbit length $l_{pt}$ is expanded with respect
to $\bm{s}_\perp=(s_2,\cdots,s_p)$ as
\begin{gather}
l_{pt}(\bm{s})=L_{pt}+\frac{1}{4R}\sum_{a,b\geq 2}
K_{ab}s_a s_b + O(s_\perp^3),
\end{gather}
where $K$ is the $(p-1)$-dimensional curvature matrix defined by
\begin{gather}
K_{ab}=2R\pp{^2l_{pt}}{s_a\partial s_b}.
\end{gather}
Integration over $s_1=R\theta$ is performed
exactly and the other integrals over $\bm{s}_\perp$ are carried out
using the SPA.  This gives
\begin{align}
&\int ds_1\cdots ds_p\, e^{ikl_{pt}(\bm{s}_\perp)} \NN
&=R\,(\theta_{\rm A'}-\theta_{\rm A})
\frac{(4\pi iR/k)^{(p-1)/2}}{\sqrt{|\det K|}}
e^{ikL_{pt}-i\pi n_-/2},
\end{align}
where $n_-$ is the number of negative eigenvalues of $K$.  In the
above procedure, only the integration range of the parameter $\theta$
is different from the case of nontruncated circular billiard.
Consequently, one has the contribution of the orbit $(p,t)$ in the
truncated circle part as
\begin{equation}
g_{pt}^{\rm (pr)}(k)=f_p g_{pt}^{\rm (circ)}(k), \label{eq:trace_pr}
\end{equation}
where $f_p$ denotes the relative volume of the parameter space
occupied by the truncated family of orbit compared with that for the
non-truncated circular billiard,
\begin{equation}
f_p=\frac{\theta_{\rm A'}-\theta_{\rm A}}{2\pi/p}
=1-\frac{p}{\pi}\,\theta_A.
\end{equation}
The term given by (\ref{eq:trace_pr}), which will be referred to as
the principal term, corresponds to the contribution of the shaded area
shown in Fig.~\ref{fig:marginal}(b).

Since the above family is truncated at $\theta=\theta_A$ and
$\theta_A'$, integration over $\theta$ in the outer region
$(\theta<\theta_{\rm A}$ and $\theta>\theta_{\rm A'})$ should also be
executed using the SPA, and it gives the end-point corrections to the
principal contribution.  In calculating the end-point corrections, the
first vertex of the orbit is supposed to be on the edge of the circle
wall, and its stability is calculated for the outer wall neighboring
the circle part.  In practice, the outer wall is extended into the
interior region ($\theta>\theta_{\rm A}$ and $\theta<\theta_{\rm A'}$)
for this vertex, as illustrated with thick broken lines $\mathrm{AP'}$
and $\mathrm{C'Q}$ in Fig.~\ref{fig:opencirc}, so that the curvature
of the surface is continuous there.  Then the contour plot of the
orbit length looks like that shown in Fig.~\ref{fig:marginal}(c), and
the orbits at the end of the family become isolated.  For these
hypothetical isolated orbits, which I call marginal orbits, it is
possible to calculate their monodromy matrices and Maslov indices in a
standard numerical prescription.  Thus, the end-point correction to
the truncated family contribution, from the shaded area shown in
Fig.~\ref{fig:marginal}(c), is given by the half of the Gutzwiller
formula (\ref{eq:Gutz_trace}) for each of those isolated marginal
orbits $\beta$ as
\begin{equation}
g_{pt,\beta}^{\rm(mg)}(k)
=2R A_{pt,\beta}^{\rm(mg)}
\sin\left(kL_{pt}-\tfrac{\pi}{2}\mu_{pt,\beta}'\right),
\label{eq:trace_mg}
\end{equation}
with the amplitude
\begin{equation}
2RA_{pt,\beta}^{\rm(mg)}
=\frac{w_{pt}L_{pt}}{2\pi\sqrt{|\det(\mathsf{M}_{pt,\beta}-I)|}}.
\end{equation}
In Eq.~(\ref{eq:trace_mg}), the sine function is taken for the
definition of the Maslov index, in contrast to the cosine one in
Eq.~(\ref{eq:Gutz_trace}), in accord with Eq.~(\ref{eq:trace_circ}).

Finally, the total contribution of the family of orbit $(p,t)$
confined in the circle part is given by the sum of the principal and
marginal terms as
\begin{equation}
g_{pt}^{\rm(tot)}(k)=g_{pt}^{\rm(pr)}(k)
+\sum_\beta g_{pt,\beta}^{\rm(mg)}(k),
\end{equation}
where the sum in the second term is taken over the two marginal orbits
$\beta$.

\subsection{Three-dimensional truncated spherical cavity}
\label{sec:3D}

\begin{figure}[tb]
\centering
\includegraphics[width=.6\linewidth]{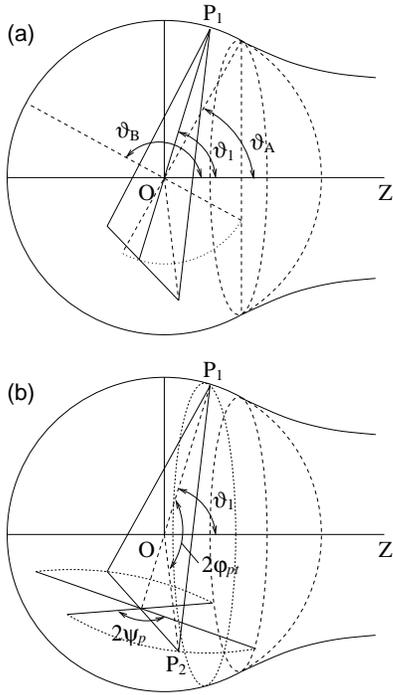}
\caption{\label{fig:opencavity}
Geometry of the periodic-orbit family confined in the 3D truncated
spherical cavity potential.  The top panel (a) shows the range of the
angle $\vartheta_1$ of the first vertex $\mathrm{P_1}$, which can vary
from $\vartheta_A$ to $\vartheta_B$.  For a given value of
$\vartheta_1$, the orbit can rotate $2\pi$ around the symmetry axis
OZ.  The orbit can be also rotated about the axis $\mathrm{OP_1}$, and
the bottom panel (b) shows the range $\psi_p$ for this rotation which
is dependent on $\vartheta_1$.}
\end{figure}

Next I consider a 3D cavity potential whose wall partly consists of a
truncated sphere with radius $R$.  As shown in
Fig.~\ref{fig:opencavity}, the spherical part is centered at O,
truncated with the plane perpendicular to the axis OZ, and smoothly
connected to the neighboring part of the surface which is assumed to
be axially symmetric about the axis OZ.  In the truncated spherical
cavity, one has the same set of periodic-orbit families as in the
complete (nontruncated) spherical cavity, but with the restricted
ranges of the parameters.

For instance, consider a family of triangle orbits confined in the
spherical part as shown in Fig.~\ref{fig:opencavity}.  The position of
the first vertex $\mathrm{P_1}$ is determined by the polar angle
$\vartheta_1$ and azimuthal angle $\varphi_1$ around the symmetry axis
OZ.  One obviously has a restriction on the polar angle,
$\vartheta_A<\vartheta_1<\vartheta_B$ as shown in
Fig.~\ref{fig:opencavity}(a), so that the entire orbit fits the
confines in the spherical part.  In general, the maximum angle
$\vartheta_B$ for the orbit $(p,t)$ is given by
\begin{equation}
\vartheta_B=\left\{
 \begin{array}{l}
  \pi-\vartheta_A \qquad \text{for even $p$,} \\
  \pi\phantom{{}-\vartheta_A} \qquad
  \text{for odd $p$ with $\vartheta_A\leq\frac{\pi}{p}$}, \\
  \cos^{-1}\left[-\dfrac{\cos\vartheta_A}{\cos(\pi/p)}\right] \quad
  \text{for odd $p$ with $\vartheta_A > \frac{\pi}{p}$}.
 \end{array}\right.
\end{equation}

\begin{figure}[tb]
\centering
\includegraphics[width=.6\linewidth]{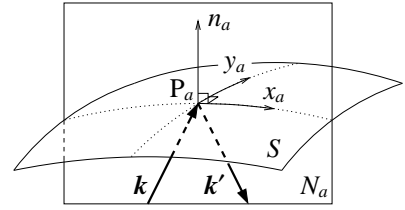}
\caption{\label{fig:local_coord}
Definition of the local coordinates around the vertex $\mathrm{P}_a$
on the surface $S$.  $\bk$ and $\bk'$ represent the wave vectors
before and after reflection at $\mathrm{P}_a$, and they define the
orbital plane $N_a$.  $(x_a,y_a)$ are local surface coordinates
tangent to $S$, where $x_a$ is taken in the orbital plane $N_a$ and
$y_a$ orthogonal to $x_a$.  $n_a$ is normal to $S$ toward exterior
region.}
\end{figure}

Let us consider the principal part of the integrals in
Eq.~(\ref{eq:trace_gen}), taking into account the degeneracies of the
orbit.  As illustrated in Fig.~\ref{fig:local_coord}, the local
surface coordinates $(x_a,y_a)$ are defined on the surface $S$ around
the vertex $\mathrm{P}_a$ of a given periodic orbit, where $x_a$ is
taken along the orbital plane $N_a$, and $y_a$ perpendicular to it.

Let us first consider a diameter family (2,1).  In this case, the
orbit in the family is uniquely specified by fixing the position of
the first vertex $\mathrm{P_1}$, and the coordinates $(x_1,y_1)$ are
varied in the available range, which fixes the orbit in the family.
Then, in Eq.~(\ref{eq:trace_gen}), the integrals over $(x_1,y_1)$ are
done exactly so as to include all members of the family, and the other
$(2p-2)$ integrals are carried out using the SPA.  The integration
over $(x_1=R\vartheta_1,y_1=R\varphi_1\sin\vartheta_1)$ gives
\begin{align}
&\int dx_1dy_1=\int Rd\vartheta_1\cdot R\sin\vartheta_1 d\varphi_1 \NN
&\quad =2\pi R^2\int_{\vartheta_A}^{\pi-\vartheta_A}
\sin\vartheta d\vartheta=4\pi R^2\cos\vartheta_A. \label{eq:int_deg2}
\end{align}
I define $f_2$ as the relative volume of the parameter space occupied
by the truncated diameter family compared with that for the complete
(nontruncated) spherical cavity, which will be called ``occupation
rate'' for brevity.  Since the same integral as (\ref{eq:int_deg2}) in
the complete spherical cavity gives the factor $4\pi R^2$, one obtains
\begin{equation}
f_2=\cos\vartheta_A.
\label{eq:fp_diam}
\end{equation}
Since the remaining $(2p-2)$ integrals using the SPA give a result
equivalent to that for the complete spherical cavity, the principal
contribution of the truncated diameter family is given by
\begin{equation}
g_{m(2,1)}^{\rm(pr)}(k)
=f_2 g_{m(2,1)}^{\rm(sph)}(k),
\end{equation}
with $g_{m(2,1)}^{\rm(sph)}(k)$ given by Eq.~(\ref{eq:trace_sph2}).

For polygon family $(p,t)$ ($p>2t$), after fixing the first vertex
$\mathrm{P_1}$, one can further rotate the orbit about the axis
$\mathrm{OP_1}$ as shown in Fig.~\ref{fig:opencavity}(b).  I define
the rotation angle $\psi$ of the orbital plane around the axis
$\mathrm{OP_1}$ measured from the position where it is perpendicular
to the plane defined by the symmetry axis OZ and the rotation axis
$\mathrm{OP_1}$.  Its maximum value $\psi_p$ is given by
\begin{equation}
\sin\psi_p(\vartheta_1)=
 \frac{\cos\vartheta_A-\cos(2\pi j/p)
 \cos\vartheta_1}{\sin(2\pi j/p)\sin\vartheta_1},
\label{eq:psi_p}
\end{equation}
if the vertex $\mathrm{P}_{j+1}$ with $\angle\mathrm{P_1OP}_{j+1}=2\pi
j/p$ first touches the joint circle by the above rotation.  If no
vertices touch the joint with the rotation, one simply has
$\psi_p=\pi/2$.  Thus, for the polygon family, the integrals in
(\ref{eq:trace_gen}) should be done exactly for three coordinates:
$x_1=R\vartheta_1$, $y_1=R\varphi_1\sin\vartheta_1$, and
$y_2=R\psi\sin(2\varphi_{pt})$.  For a given $(x_1,y_1)$, the integral
over $\psi$ simply gives $4\psi_p(\vartheta_1)$, and one obtains
\begin{gather}
\int dx_1 dy_1 dy_2
=\int Rd\vartheta_1~ R\sin\vartheta_1 d\varphi_1~
R\sin(2\varphi_{pt})d\psi \NN
=8\pi R^3\sin(2\varphi_{pt})\int_{\vartheta_A}^{\vartheta_B}
\psi_p(\vartheta)\sin\vartheta d\vartheta.
\label{eq:int_degp}
\end{gather}
Since the same integral for the complete spherical cavity gives $8\pi
R^3\sin(2\varphi_{pt})$, one obtains the occupation rate $f_p$ for the
family $(p,t)$ as
\begin{equation}
f_p=\int_{\vartheta_A}^{\vartheta_B}\psi_p(\vartheta)
\sin\vartheta d\vartheta. \label{eq:fp}
\end{equation}
The remaining $(2p-3)$ integrations in (\ref{eq:trace_gen}) using the
SPA give exactly the same results as those for the complete spherical
cavity, and the principal contribution of the polygon family $(p,t)$
to the level density is expressed as
\begin{equation}
g_{pt}^{\rm(pr)}(k)=f_p g_{pt}^{\rm(sph)}(k)
\end{equation}
with $g_{pt}^{\rm(sph)}(k)$ given by Eq.~(\ref{eq:trace_sphere}).

\begin{figure}[tb]
\centering
\includegraphics[width=\linewidth]{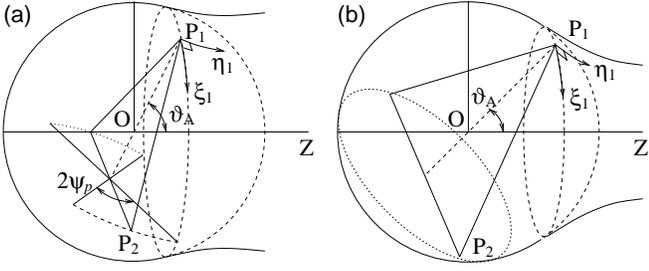}
\caption{\label{fig:local_coord2}
Local coordinates for a marginal orbit with its vertex $\mathrm{P_1}$
on the joint.  Around the vertex $\mathrm{P_1}$, local coordinate
$\xi_1$ is taken along the joint circle and $\eta_1$ perpendicular to
it.  For $\vartheta_A$ close to $\pi/2$, the vertex $\mathrm{P_2}$
touches the joint circle by rotating the orbit about the axis
$\mathrm{OP_1}$ [panel (a)], but it can rotate $2\pi$ around the axis
$\mathrm{OP_1}$ for sufficiently small $\vartheta_A$ [panel (b)].}
\end{figure}

In addition to the above principal terms, one should consider the
end-point corrections.  They are associated with the marginal orbits
whose first vertex $\mathrm{P_1}$ is on the joint of the spherical and
the neighboring walls as shown in Fig.~\ref{fig:local_coord2}.  For
such orbits, the local surface coordinate $(x_1,y_1)$ is transformed
into $(\xi_1,\eta_1)$ so that $\xi_1$ is along the joint circle and
$\eta_1$ perpendicular to it along the surface outside the spherical
wall, as illustrated in Fig.~\ref{fig:local_coord2}.  Then, the
integral over $\eta_1>0$ is executed using the SPA.  The contribution
of the marginal orbit is obtained, just in the same manner as the
billiard case, by extending the wall neighboring the spherical part
into the inner region around $\mathrm{P_1}$ so that the curvature of
the surface is continuous there.  It makes the marginal orbits a
family with reduced degeneracy whose symmetry-reduced monodromy
matrices and Maslov indices can be calculated in the standard
prescription.  For the marginal diameter family, which forms a
one-parameter family generated by the rotation about the symmetry
axis, the integral over $\xi_1$ is executed exactly and the rest of
the $(2p-1)$ integrals are carried out using the SPA.  The
contribution to the level density is expressed in the form
\begin{equation}
g_{21}^{\rm(mg)}(k)=2R\sqrt{kR}A_{21}^{\rm(mg)}
\sin\left(kL_{21}-\tfrac{\pi}{2}\mu_{21}'\right),
\end{equation}
where $A_{21}^{\rm(mg)}$ represents the dimensionless
energy-independent amplitude factor.  For the marginal polygon family,
which forms a two-parameter family, the integrals with respect to
$(\xi_1,y_j)$ are executed exactly and the rest of the $2p$ integrals
are carried out by using the SPA.  The result is expressed in the form
\begin{equation}
g_{pt}^{\rm(mg)}(k)=2R(kR)A_{pt}^{\rm(mg)}
\sin\left(kL_{pt}-\tfrac{\pi}{2}\mu_{pt}'\right).
\end{equation}
See the Appendix~\ref{sec:appA} for the expressions of the amplitude
factors $A_{pt}^{\rm (mg)}$.

For a marginal polygon family for which Eq.~(\ref{eq:psi_p}) gives the
angle $\psi_p(\vartheta_A)<\pi/2$, one should consider a secondary
marginal orbit which has two vertices $\mathrm{P_1}$ and
$\mathrm{P}_j$ on the joint.  It forms a one-parameter family, and its
contribution is evaluated by executing the integral over $\xi_1$
exactly and the other $(2p-1)$ integrals through the SPA to obtain
\begin{equation}
g_{pt}^{\rm(mm)}(k)=2R\sqrt{kR}A_{pt}^{\rm(mm)}
\sin\left(kL_{pt}-\tfrac{\pi}{2}\mu_{pt}''\right).
\end{equation}
The expression of the amplitude factor $A_{pt}^{\rm(mm)}$ is also
given in the Appendix~\ref{sec:appA}.

Finally, the total contribution of the family of orbit $(p,t)$
confined in the spherical part of the wall is given by
\begin{equation}
g_{pt}^{\rm(tot)}(k)
=g_{pt}^{\rm(pr)}(k)+g_{pt}^{\rm(mg)}(k)+g_{pt}^{\rm(mm)}(k).
\end{equation}

\section{Applications to the three-quadratic-surfaces potentials}
\label{sec:applications}

\subsection{The three-quadratic-surfaces parametrization}

As the applications of the trace formula obtained above, I consider
the 2D billiard as well as the axially symmetric 3D cavity with the
three-quadratic-surfaces (TQS) parametrization\cite{Nix69}, which is
designed to describe nuclear fission processes.  The cavity potential
is known to preserve the important characteristics of single-particle
shell structures in more realistic nuclear mean-field models such as
the Woods-Saxon potential with spin-orbit coupling\cite{StrMag77}.
Thus, the results obtained with this model will help one understand
some essential features of the fission dynamics at least
qualitatively.

\begin{figure}[tb]
\centering
\includegraphics[width=.7\linewidth]{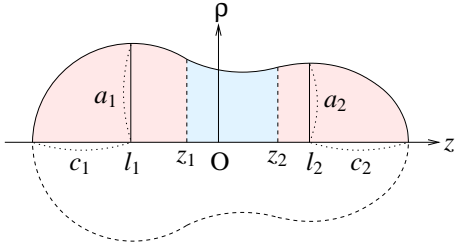}
\caption{\label{fig:tqs_params}
TQS parametrization for the shape of the nuclear fission process;
see Eq.~(\ref{eq:tqs_formula}) for the parameters.}
\end{figure}

\begin{figure}[tb]
\centering
\includegraphics[width=\linewidth]{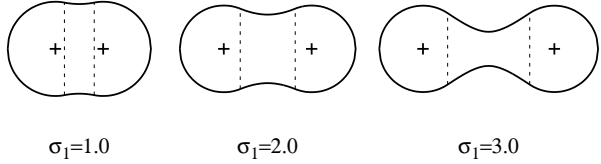}
\caption{\label{fig:tqs_wall}
Shapes of the symmetric TQS wall with $\sigma_2=-0.6$, $\sigma_3=1$,
and several values of the elongation parameter $\sigma_1$.}
\end{figure}

The potential wall consists of the two prefragment parts and the neck
part between them as shown in Fig.~\ref{fig:tqs_params}, and each of
those three parts is given by the axially symmetric quadratic surface
$\rho=\rho_s(z)$ expressed as
\begin{gather}
\rho_s^2(z)=\left\{\begin{array}{l@{\qquad}l}
a_1^2-\frac{a_1^2}{c_1^2}(z-l_1)^2 & (z_{\rm min}<z<z_1) \\
a_3^2-\frac{a_3^2}{c_3^2}(z-l_3)^2 & (z_1<z<z_2) \\
a_2^2-\frac{a_2^2}{c_2^2}(z-l_2)^2 & (z_2<z<z_{\rm max})
                   \end{array}\right.
\label{eq:tqs_formula}
\end{gather}
On the right-hand side, the first and third lines describe the left
and right prefragments centered at $z=l_1$ and $l_2$.  The second line
describes the neck part, which is smoothly connected to the left and
right prefragments at $z=z_1$ and $z_2$, respectively.
Note that the neck part is concave in most cases, for which one has
$c_3^2<0$.  The continuity of $\rho_s(z)$ and $\rho_s'(z)$ at the
joints $z=z_1$ and $z_2$ impose four constraints on the eleven
parameters $\{a_{1-3},c_{1-3},l_{1-3},z_{1,2}\}$, and the
center-of-mass and volume conservation conditions impose two more
constraints.  (The last two conditions are imposed assuming the 3D
axial cavity, and I use the same values of the parameters also in the
2D billiard.)  Thus, one eventually has five free parameters which
describe the shape of the potential.  Among the useful choices for the
shape parameters\cite{Nix69} are $\{\sigma_{1-3},\alpha_{1-3}\}$
defined by
\begin{gather}
\left\{\begin{array}{l}
\sigma_1=\frac{l_2-l_1}{u}, \quad
\alpha_1=\frac{l_1+l_2}{2u} \quad
 \text{with} \quad u=\sqrt{\frac{a_1^2+a_2^2}{2}}, \\
\sigma_2=\frac{a_3^2}{c_3^2}, \quad
\alpha_2=\frac{a_1^2-a_2^2}{u^2}, \\
\sigma_3=\frac12\left(\frac{a_1^2}{c_1^2}+\frac{a_2^2}{c_2^2}
       \right), \quad
\alpha_3=\frac{a_1^2}{c_1^2}-\frac{a_2^2}{c_2^2}.
       \end{array}\right.
\label{eq:tqs_params}
\end{gather}
The parameter $\sigma_1$ describes the elongation, $\sigma_2$ gives
the neck curvature, $\alpha_2$ is related to the fragment-mass
asymmetry, and $\sigma_3$ and $\alpha_3$ determine the shapes of the
prefragments.  The parameter $\alpha_1$ represents the asymmetry of
the positions of the prefragments from the center of mass, and it is
automatically determined by the other 5 parameters.  In this paper, I
only consider the case of symmetric shapes $(\alpha_{1-3}=0)$ with
fixed neck curvature $(\sigma_2=-0.6)$ and spherical prefragments
($\sigma_3=1$).  The shapes of the wall for several values of
$\sigma_1$ are shown in Fig.~\ref{fig:tqs_wall}.  Results for
asymmetric shapes, which are significant in investigating the origin
of asymmetric fission, will be presented in a separate
paper\cite{AIM2018}.

\subsection{Classical periodic orbits in the TQS wall}

I discuss here the properties of classical periodic orbits in the TQS
billiard and cavity potentials.  Let us first consider the 2D billiard
system.  Since the TQS wall under consideration consists partly of
circular arcs, one has degenerate family of orbits confined in each of
them.  The radius of the circle is $a_1=c_1\equiv R$ and the angle
$\theta_A$ in Fig.~\ref{fig:opencirc} for the TQS model is given by
\begin{gather}
\cos\theta_A=\frac{z_1-l_1}{R}.
\end{gather}
The diameter orbit is the only degenerate family confined in the
prefragment part in the deformation range $0\leq\sigma_1\leq 2.5$, for
which $\theta_A$ is always greater than $\pi/3$.  The triangle family
appears at $\sigma_1\simeq 2.67$ and the square family at
$\sigma_1\simeq 3.78$.

\begin{figure}[tb]
\centering
\includegraphics[width=\linewidth]{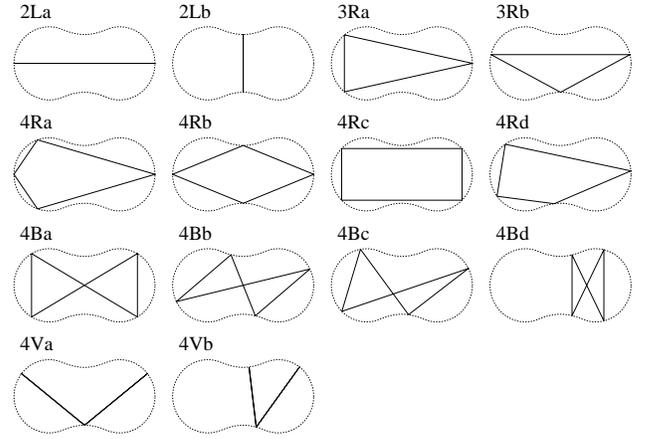}
\caption{\label{fig:orbits}
Some short isolated periodic orbits in the symmetric TQS billiard
potential for $\sigma_1=2.0$, $\sigma_2=-0.6$ and $\sigma_3=1$.  The
name of the orbit is given after the number of vertices, the type of
the shape (abbreviation of linear, rotational, butterfly, or V
shaped), and an alphabetic identifier.}
\end{figure}

Besides the degenerate family in the prefragments, there are isolated
orbits which hit the neck part of the wall or go back and forth
between two prefragments.  Some of those isolated orbits for
$\sigma_1=2.0$ are shown in Fig.~\ref{fig:orbits}.
Since the neck part has negative curvature, the orbits reflected on
the neck surface are strongly unstable (chaotic) and are expected to
make only a small contribution to the level density.
\begin{figure}[tb]
\centering
\includegraphics[width=\linewidth]{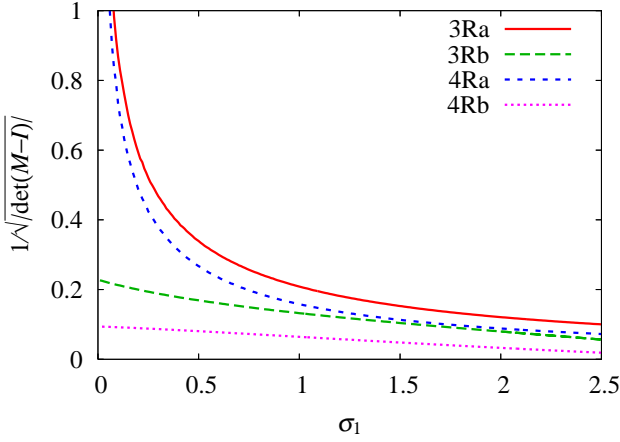}
\caption{\label{fig:stability}
Stability factor $1/\sqrt{|\det(\mathsf{M}_\beta-I)|}$ for several
isolated periodic orbits shown in Fig.~\ref{fig:orbits}.}
\end{figure}
Figure~\ref{fig:stability} shows the stability factor
$1/\sqrt{|\det(\mathsf{M}_\beta-I)|}$ in the Gutzwiller trace formula
(\ref{eq:Gutz_trace}) for some short isolated orbits shown in
Fig.~\ref{fig:orbits}.  For the orbits 3Rb and 4Rb, which have
vertices on the neck part, the value of the stability factor is
considerably smaller than those for the orbits 3Ra and 4Ra, which have
vertices only on the prefragment parts.  One sees that the latter also
become chaotic with increasing elongation parameter $\sigma_1$.

In the 3D TQS cavity potential, one has families of all the regular
polygons $(p,t)$ confined in the spherical prefragment parts.  As in
the nontruncated spherical cavity, the polygon orbits $(p>2t)$ form
three-parameter families and the diameter orbits $(p=2t)$ form
two-parameter families.  The orbits shown in Fig.~\ref{fig:orbits} in
this case will form a one-parameter family generated by the rotation
about the symmetry axis, except the diameter 2La which remains
isolated.  In addition, one has orbits on the equatorial plane in the
neck surface and three-dimensional orbits, which also form
one-parameter families.  From the viewpoint of semiclassical expansion
with respect to the degeneracy, the shell effect might be mainly
governed by the three-parameter polygon families in the prefragments,
with relatively small contribution of the two-parameter diameter
family, and the other one-parameter families might play only minor
roles.

\subsection{Fourier analysis}

\begin{figure}[tb]
\centering
\includegraphics[width=\linewidth]{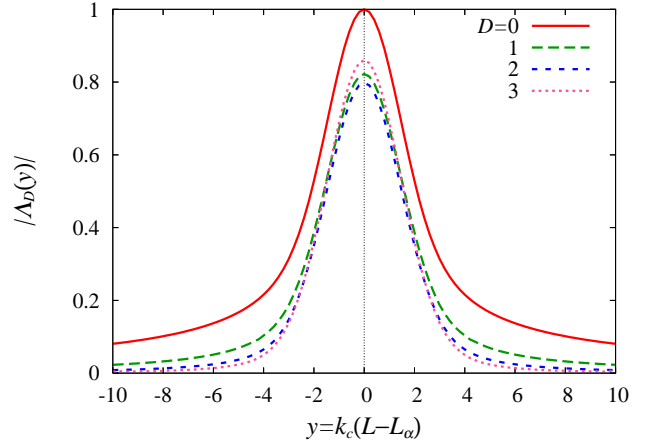}
\caption{\label{fig:fshape}
Modulus of the function $\varLambda_D(y)$ defined by
Eq.~(\ref{eq:fshape}) for several values of $D$.}
\end{figure}

In the billiard and cavity systems, the action integral along the
orbit is given by a simple product of the wave number $k$ and the
orbit length $L_\beta$.  Owing to such a simple energy dependence of
the phase part, one obtains a clear correspondence between the
classical periodic orbits and quantum level density through the
Fourier analyses.  The Fourier transform of the level density defined
by
\begin{equation}
F(L)=\sqrt{\frac{2}{\pi}}\int_0^\infty dk\,g(k)e^{ikL}e^{-(k/k_c)^2/2},
\label{eq:fourier}
\end{equation}
is considered, where a Gaussian factor with the cutoff momentum $k_c$
is introduced in the integrand to truncate the high energy part $k\gg
k_c$ of the level density unavailable in the numerical calculation.
Inserting the quantum level density $g(k)=\sum_i\delta(k-k_i)$, one
has
\begin{equation}
F^{\rm(qm)}(L)=\sqrt{\frac{2}{\pi}}\sum_i e^{ik_iL}e^{-(k_i/k_c)^2/2}
\label{eq:fourier_qm}
\end{equation}
which can be easily evaluated using the quantum spectrum.  The
semiclassical level density in a hard-wall potential model (either
in 2D or 3D) is generally expressed in a form
\begin{equation}
g(k)=g_0(k)+2R_0\sum_\beta (kR_0)^{D_\beta/2}A_\beta
\sin\left(kL_\beta-\tfrac{\pi}{2}\mu_\beta\right),
\label{eq:g_scdeg}
\end{equation}
where $D_\beta$ is the degeneracy of the orbit family $\beta$.
Inserting (\ref{eq:g_scdeg}) into (\ref{eq:fourier}), one has
\begin{align}
F^{\rm(sc)}(L)=F_0(L)+i\sum_\beta
& (k_cR_0)^{1+D_\beta/2}A_\beta e^{i\pi\mu_\beta/2} \nonumber \\
& \times \varLambda_{D_\beta}(k_c(L-L_\beta)),
\label{eq:fourier_sc}
\end{align}
where
\begin{equation}
\varLambda_D(y)=\sqrt{\frac{2}{\pi}}\int_0^\infty
 dx\,x^{D/2}e^{iyx}e^{-x^2/2} \label{eq:fshape}
\end{equation}
is a function whose modulus has a peak at the origin $y=0$ as
shown in Fig.~\ref{fig:fshape}.
Thus, the Fourier amplitude of the quantum level density calculated
with Eq.~(\ref{eq:fourier_qm}) will exhibit successive peaks at the
lengths of classical periodic orbits, whose heights are proportional
to the amplitude factor $A_\beta$ of the semiclassical level density.
If the cutoff momentum $k_c$ is large enough to single out the peak of
the orbit $\beta$ from those of the other orbits, the modulus of the
Fourier transform at $L=L_\beta$ is given by
\begin{gather}
|F(L_\beta)|=(k_cR_0)^{1+D_\beta/2}\varLambda_{D_\beta}(0)
A_\beta, \\
\varLambda_D(0)=\frac{2^{D/2}\varGamma(1+D/2)}{\sqrt{\pi}}.
\end{gather}
Taking account of all the marginal families, the contribution of the
orbit $(p,t)$ to the level density of the truncated circular billiard
or the truncated spherical cavity is written as
\begin{equation}
g_{pt}(k)=2R\sum_D (kR)^{D/2}A_{pt}^{(D)}
\sin\left(kL_{pt}-\tfrac{\pi}{2}\mu_{pt}^{(D)}\right)
\end{equation}
where the sum is taken over the degeneracy parameters $D$ which
differ in principal and marginal terms
(see the Appendix).  It can be written in a more compact
way as
\begin{equation}
g_{pt}(k)=2R|\mathcal{A}_{pt}(kR)|\sin\left(kL_{pt}-\tfrac{\pi}{2}
\mu_{pt}^{\rm(eff)}\right),
\end{equation}
with the complex amplitude $\mathcal{A}_{pt}$ and effective Maslov
index $\mu_{pt}^{\rm(eff)}$ defined by
\begin{gather}
\mathcal{A}_{pt}(x)=\sum_D x^{D/2} A_{pt}^{(D)}
e^{-i\pi\mu_{pt}^{(D)}/2},\\
-\frac{\pi}{2}\mu_{pt}^{\rm(eff)}(x)=\arg\mathcal{A}_{pt}(x).
\end{gather}
The Fourier transform at $L=L_{pt}$ is then given by
\begin{align}
F(L_{pt})&= i\sum_D(k_cR)^{1+D/2}
e^{i\pi\mu_{pt}^{(D)}/2}\varLambda_D(0)A_{pt}^{(D)} \\
&\approx 0.8ik_cR\mathcal{A}_{pt}^*(k_cR).
\label{eq:fourier_len}
\end{align}
In the last approximation, the value $\varLambda_{D\geq 1}(0)\approx
0.8$ is used, which can be seen in Fig.~\ref{fig:fshape}.  Thus, the
Fourier transform of the quantum level density provides us with direct
information on the amplitude $\mathcal{A}_{pt}$, which represents the
combined contribution of the principal and marginal terms.

\subsection{Two-dimensional TQS billiard}

\begin{figure}[tb]
\centering
\includegraphics[width=\linewidth]{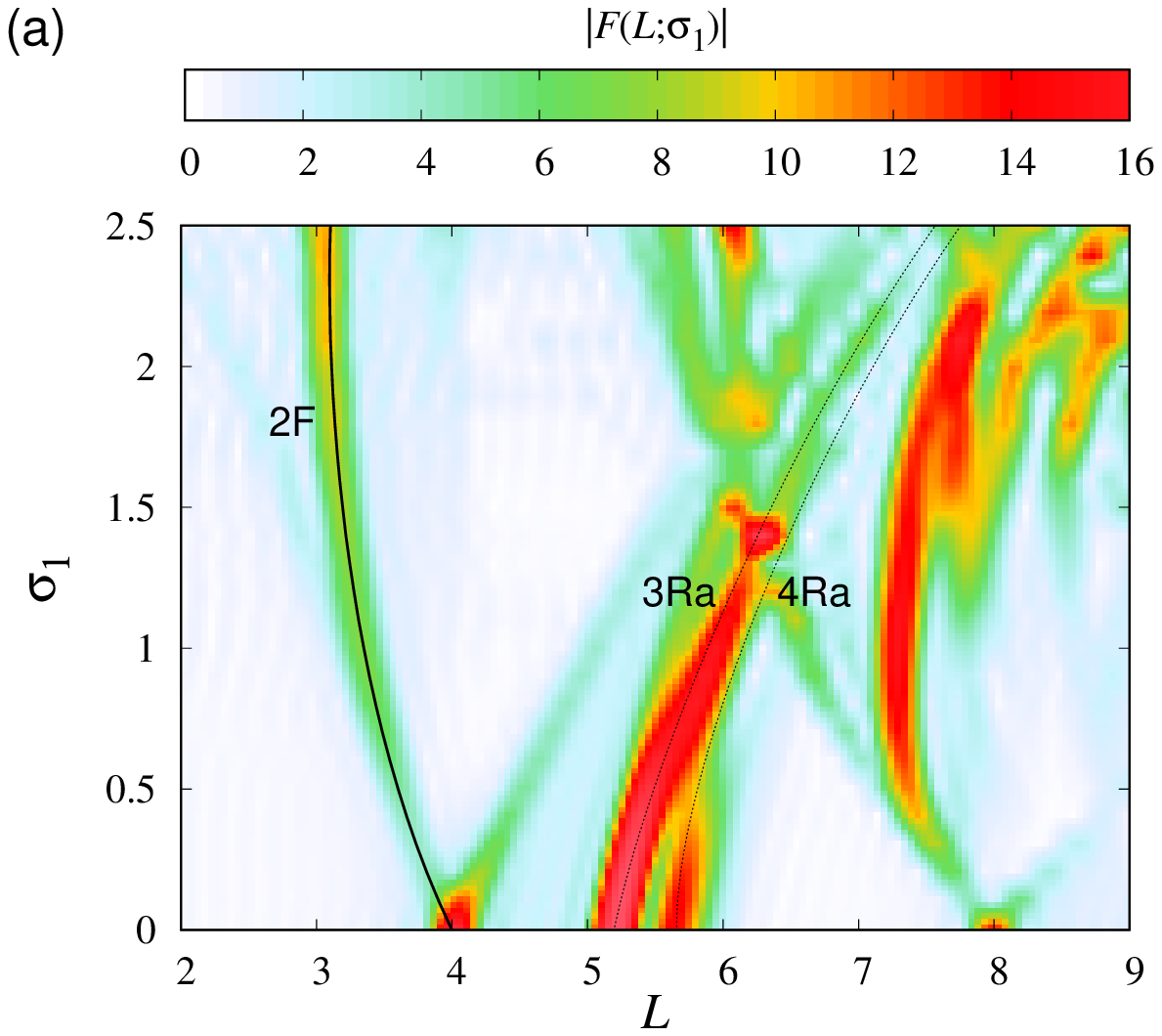} \\
\includegraphics[width=\linewidth]{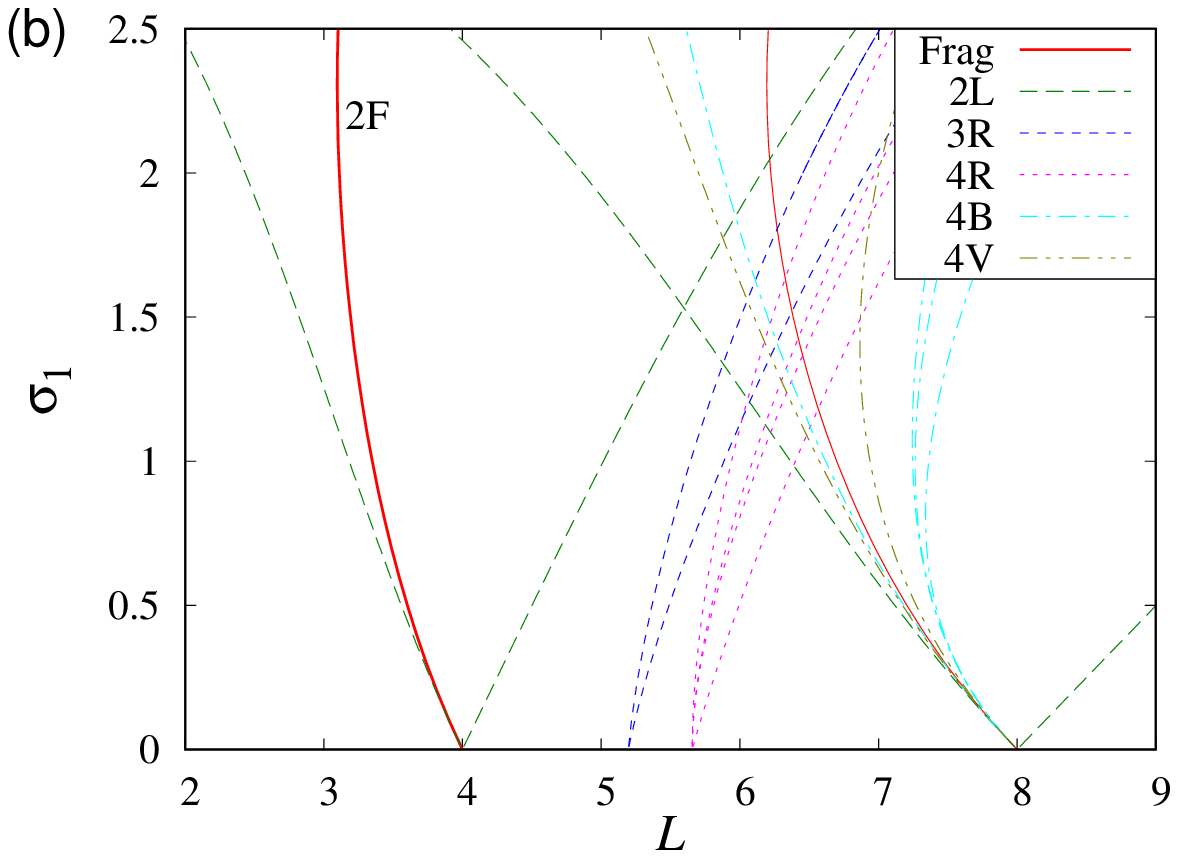}
\caption{\label{fig:fmap2d} 
The top panel (a) shows the Fourier amplitude of the quantum level
density as a function of $L$ (abscissa) and $\sigma_1$ (ordinate).  In
the bottom panel, lengths of the classical periodic orbits are plotted
as functions of $\sigma_1$ in the same range of $(\sigma_1,L)$ as the
top panel.  The solid line labeled 2F represents the diameter orbit
family confined in the prefragments.  Broken lines represent the
isolated orbits as shown in Fig.~\ref{fig:orbits}.}
\end{figure}

In this section, the quantum-classical correspondence
in the 2D TQS billiard system is investigated
using the Fourier transformation
technique discussed above.  In the top panel of Fig.~\ref{fig:fmap2d},
moduli of the quantum Fourier transform $|F^{\rm(qm)}(L;\sigma_1)|$ are
displayed as functions of $L$ and $\sigma_1$.  In the bottom
panel, lengths of some classical periodic orbits are plotted as
functions of $\sigma_1$.  As expected from the semiclassical trace
formula, Fourier amplitudes of the quantum level density show peaks at
the lengths of the classical periodic orbits.  For the circular shape
($\sigma_1=0$), one sees strong Fourier peaks at $L=L_{21}(=4)$,
$L_{31}(=3\sqrt{3}\simeq 5.19)$, and $L_{41}(=4\sqrt{2}\simeq 5.66)$
corresponding to the diameter, triangle, and square orbits,
respectively.  Those peaks promptly decay with increasing $\sigma_1$,
but the peak of the prefragment diameter orbits (labeled 2F) grows
again at large $\sigma_1$.

\begin{figure}[tb]
\centering
\includegraphics[width=\linewidth]{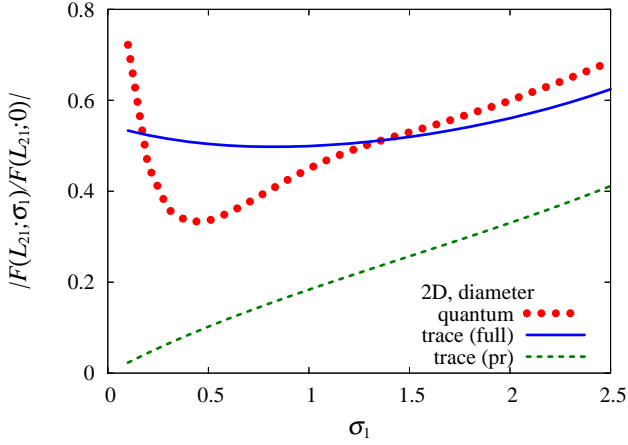}
\caption{\label{fig:ftlobt2d} 
Fourier amplitudes of the level density for the TQS billiard evaluated
at the length of the diameter orbit confined in the circle parts.  Its
value relative to that for the spherical shape is plotted as a
function of $\sigma_1$.  The thick dotted (red) line represents the
quantum result, and the solid (blue) line represents the result of the
semiclassical trace formula, taking account of the marginal orbit
contributions.  The dashed (green) line shows the semiclassical result
only with the contribution of the principal family.}
\end{figure}

Figure~\ref{fig:ftlobt2d} shows the Fourier amplitude
$|F(L_{21}(\sigma_1))|$ evaluated at the length of the diameter orbit
as functions of the deformation parameter $\sigma_1$.  The quantum
mechanical result is compared with the semiclassical one given by
(\ref{eq:fourier_len}).  It is found that the principal term
considerably underestimates the quantum result in the energy region
considered, especially for the case of small $\sigma_1$.  After taking
into account the contributions of the marginal orbits, the quantum
result is reasonably reproduced for $\sigma_1\gtrsim 1.0$.  For
smaller $\sigma_1$, the breaking of the total rotational symmetry
should be treated appropriately using a kind of uniform approximation,
but it is beyond the scope of the current work.

\begin{figure}
\centering
\includegraphics[width=\linewidth]{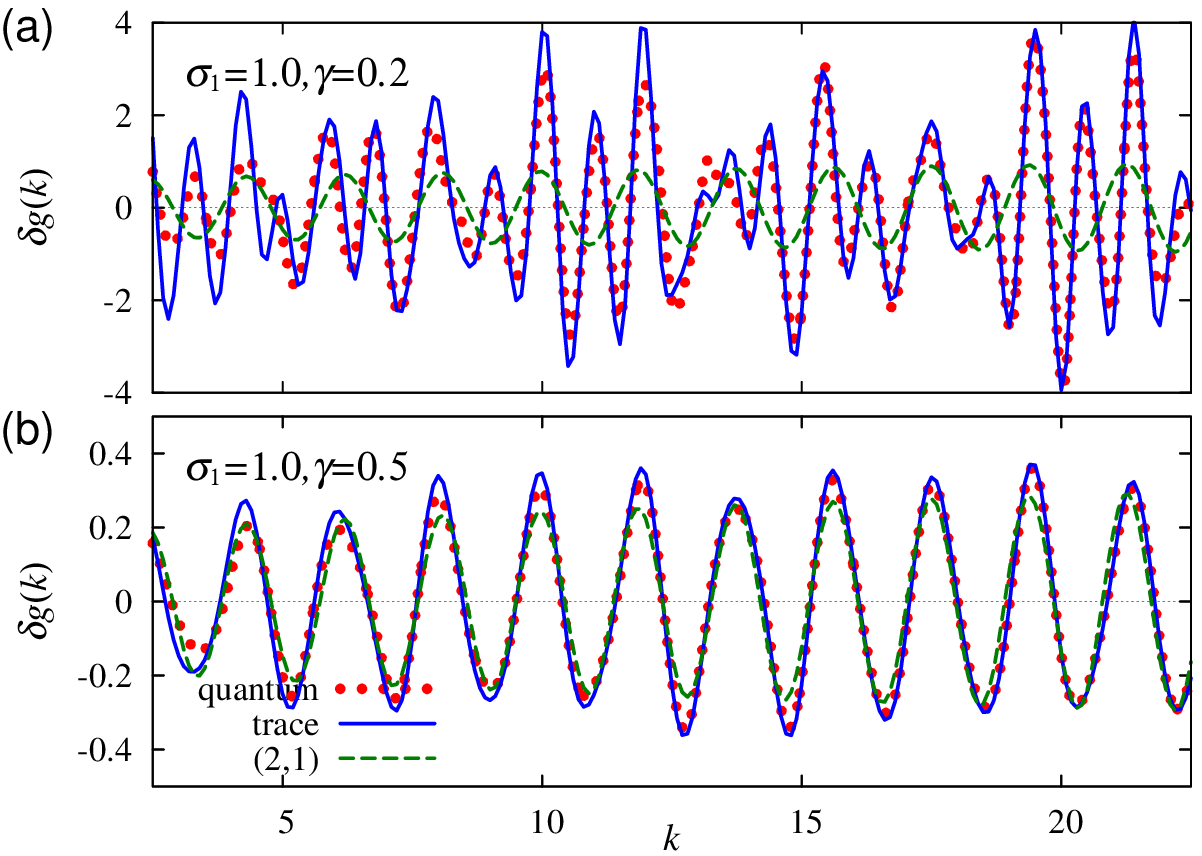} \\
\includegraphics[width=\linewidth]{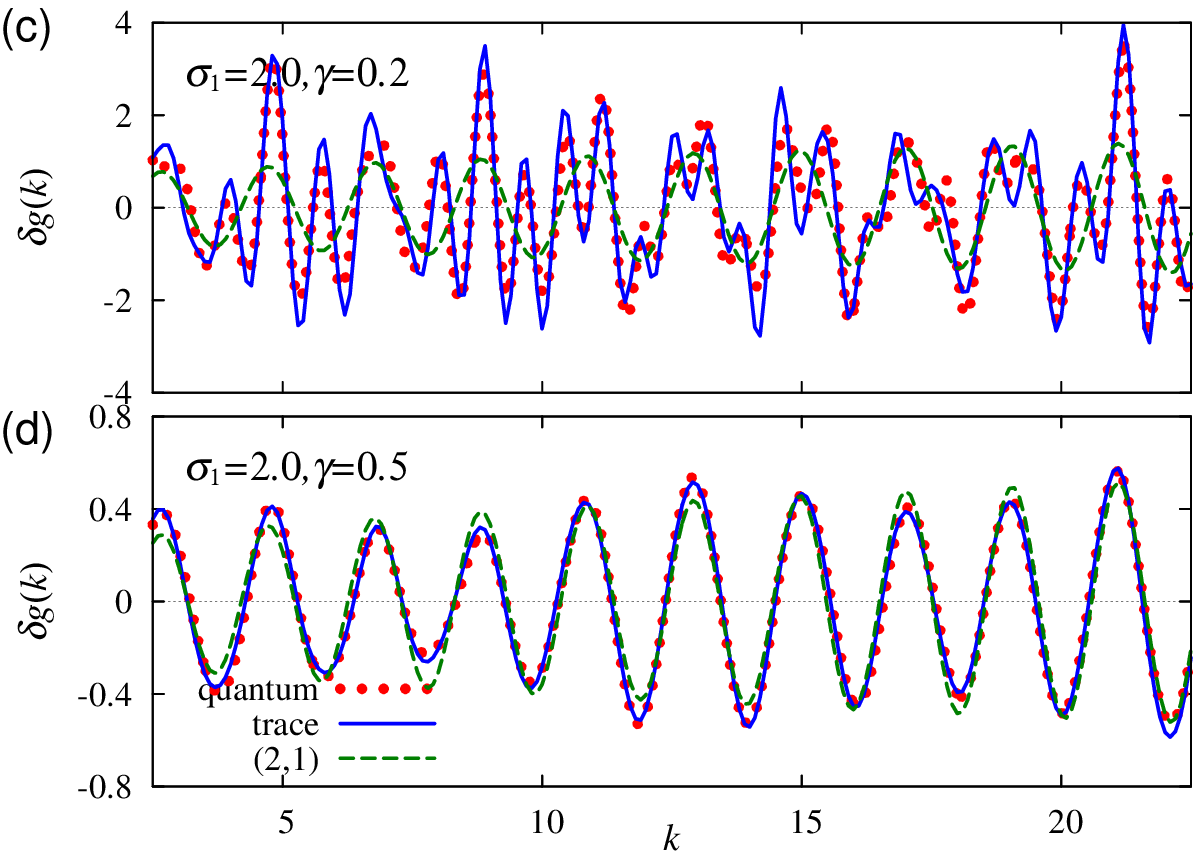}
\caption{\label{fig:sld2d} 
Oscillating part of the level density with averaging width
$\gamma=0.2$ and $0.5$ for $\sigma_1=1.0$ [panels (a) and (b)]
and 2.0 [panels (c) and (d)].  Thick dotted (red) lines represent the
quantum results, solid (blue) lines represent the trace formula, and
broken (green) lines show the contribution of the prefragment
diameter orbit family (2,1).}
\end{figure}

Figure~\ref{fig:sld2d} shows the oscillating part of the quantum and
semiclassical level densities (\ref{eq:sld_gam}) for TQS billiard with
the deformation parameter $\sigma_1=1.0$ and $2.0$.  Values of the
averaging width $\gamma=0.2$ and $0.5$ are used to see the fine and
gross shell structures, respectively.  In each panel, one sees that
the quantum results are nicely reproduced by the semiclassical trace
formulas taking account of the contribution of short isolated orbits
shown in Fig.~\ref{fig:orbits} as well as the prefragment diameter
families.  The contribution of the isolated orbits are calculated with
the Gutzwiller formula (\ref{eq:Gutz_trace}).  In the contribution of
the prefragment diameter families, both principal and marginal terms
are taken into account.  As shown in the plots for $\gamma=0.5$, the
contribution of the primitive diameter orbit family dominates the
gross shell structures.

\subsection{Three-dimensional TQS cavity}

Next, let us consider the 3D TQS cavity systems.  Unlike the 2D billiard,
one has all $(p,t)$ families confined in the spherical prefragments
for all values of $\sigma_1>0$.  The principal
contributions of the three-parameter polygon families $(p>2t)$ and
two-parameter diameter families $(p=2t)$
are obtained with the occupation
rate $f_p$ given by Eqs.~(\ref{eq:fp}) and (\ref{eq:fp_diam}) as
\begin{equation}
g_{pt}^{\rm(pr)}(k;\sigma_1)=2\sum_{pt}f_p g_{pt}^{\rm(sph)}(k;R(\sigma_1)),
\end{equation}
where $g_{pt}^{\rm(sph)}(k;R)$ represents the contribution of the
orbit family $(p,t)$ in the spherical cavity with radius $R$, and the
overall factor 2 counts the families in two prefragments which are
equivalent for the symmetric shapes.  Numerical results of the
occupation rate $f_p$ for the symmetric TQS cavity with
$\sigma_2=-0.6,~ \sigma_3=1,~ \alpha_{1-3}=0$ and several values of
$\sigma_1$ are shown in Table~\ref{table1}.
\begin{table}[b]
\centering
\caption{\label{table1}
Occupation rate (\ref{eq:fp_diam}) and (\ref{eq:fp}) for the
prefragment orbit families in the symmetric TQS cavity
with the parameters $\sigma_2=-0.6$, $\sigma_3=1$ and several values of
$\sigma_1$.  $\vartheta_A$ is the angle indicated in
Fig.~\ref{fig:opencavity}.}
\begin{tabular}{c|c|c|c|c|c} \hline
$\sigma_1$ & 1.0 & 1.5 & 2.0 & 2.5 & 3.0 \\ \hline
$\vartheta_A$ [deg] & 79.193 & 73.665 & 67.975 & 62.047 & 55.771 \\ \hline
$f_2$ & 0.18760 & 0.28088 & 0.37562 & 0.46851 & 0.56190 \\
$f_3$ & 0.02962 & 0.06823 & 0.12650 & 0.21112 & 0.34345 \\
$f_4$ & 0.02266 & 0.05169 & 0.09436 & 0.15204 & 0.22924 \\
$f_5$ & 0.02055 & 0.04683 & 0.08504 & 0.13652 & 0.20416 \\ \hline
\end{tabular}
\end{table}
With increasing $\sigma_1$, the angle $\vartheta_A$ indicated in
Fig.~\ref{fig:opencavity} becomes smaller.  Then the occupation rates
$f_p$ increase, making the contribution of the prefragment orbit
family more important.

\begin{figure}
\centering
\includegraphics[width=\linewidth]{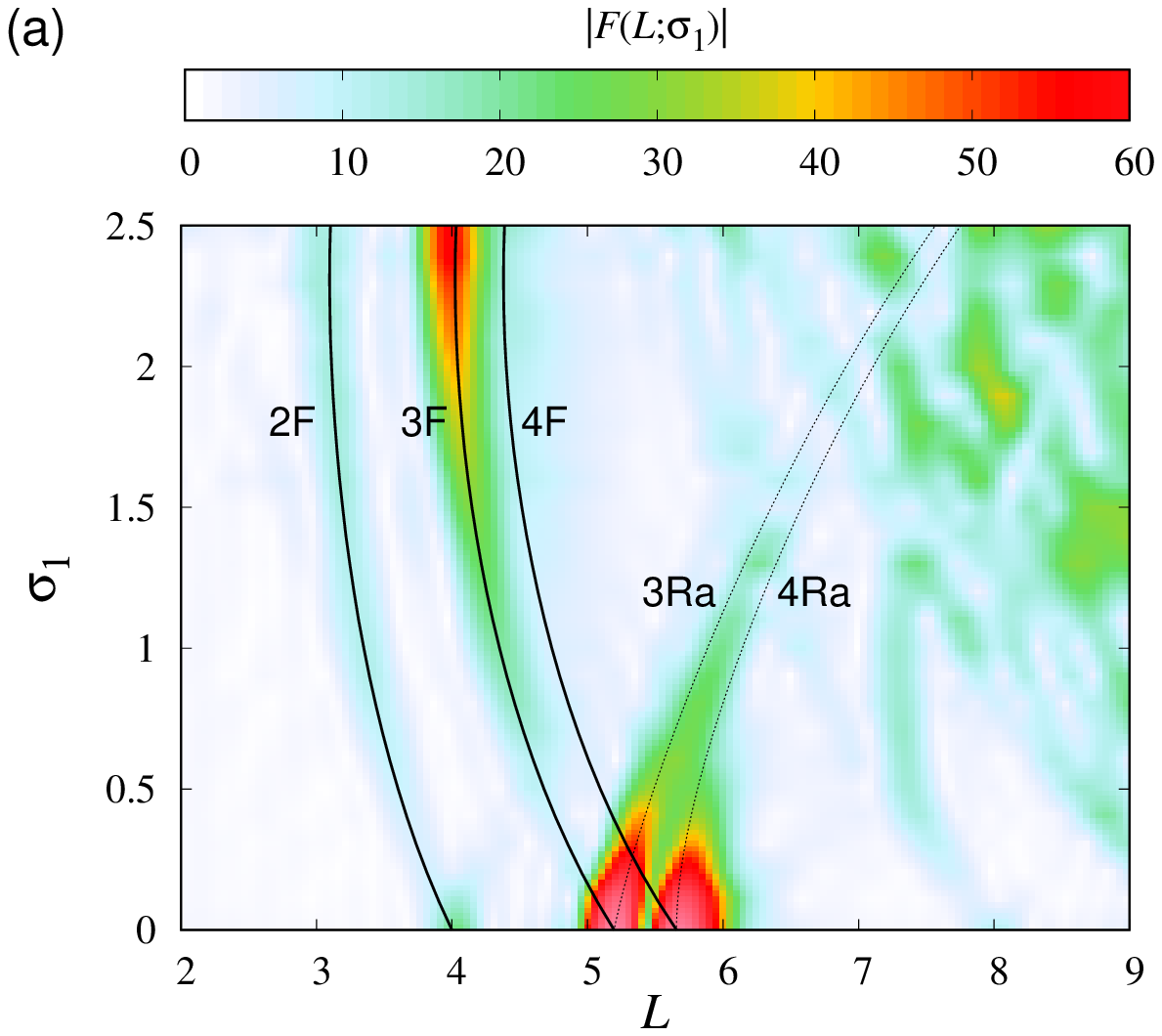} \\
\includegraphics[width=\linewidth]{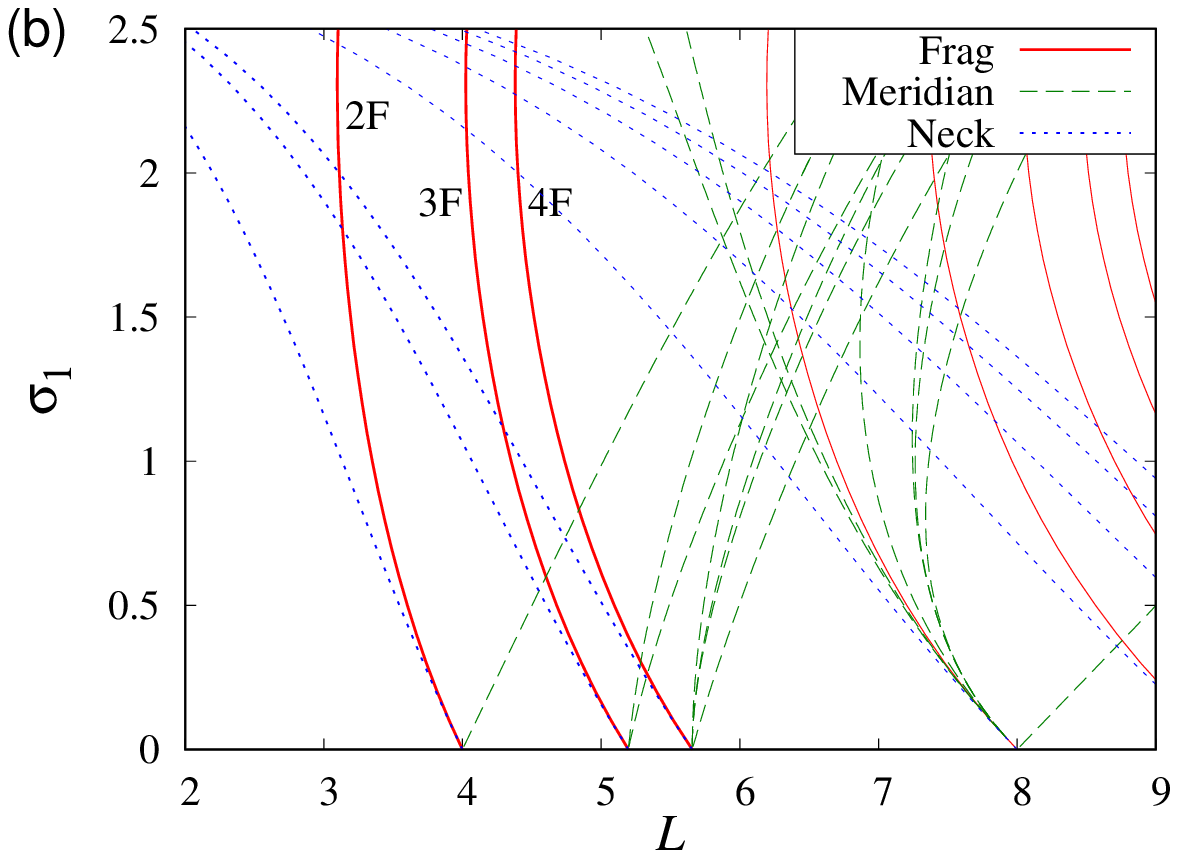}
\caption{\label{fig:fmap} 
Same as Fig.~\ref{fig:fmap2d} but for the three-dimensional TQS cavity
potential model.  Solid lines labeled 2F, 3F and 4F represent the
lengths of prefragment diameter, triangle, and square orbits,
respectively.  Broken lines represent the meridian-plane orbits as
shown in Fig.~\ref{fig:orbits}.  Dotted lines represent the equatorial
orbits in the neck part.}
\end{figure}

To quantify the contributions of the periodic orbits, let us examine
the Fourier transform of the level density (\ref{eq:fourier}) just as
in the 2D billiard case.  In Fig.~\ref{fig:fmap}, the upper panel
shows the Fourier amplitude $|F(L;\sigma_1)|$ of the quantum level
density as a function of $L$ and $\sigma_1$, and the lower panel shows
the lengths of the classical periodic orbits $L_\beta$ as functions of
$\sigma_1$ in the same ranges of $(L,\sigma_1)$ as the upper panel.
At the spherical shape $(\sigma_1=0)$, one finds especially large
peaks corresponding to the triangle and square orbits, and the peak of
the diameter orbit is relatively small due to the lower degeneracy
[note the factor $(k_cR_0)^{D_\beta/2}$ in Eq.~(\ref{eq:fourier_sc})].
With increasing $\sigma_1$, those peaks promptly decay, and then the
peaks corresponding to the prefragment orbit families begin to grow
up.  Especially, one sees a remarkable peak at the triangle family
(labeled 3F).

\begin{figure}
\centering
\includegraphics[width=\linewidth]{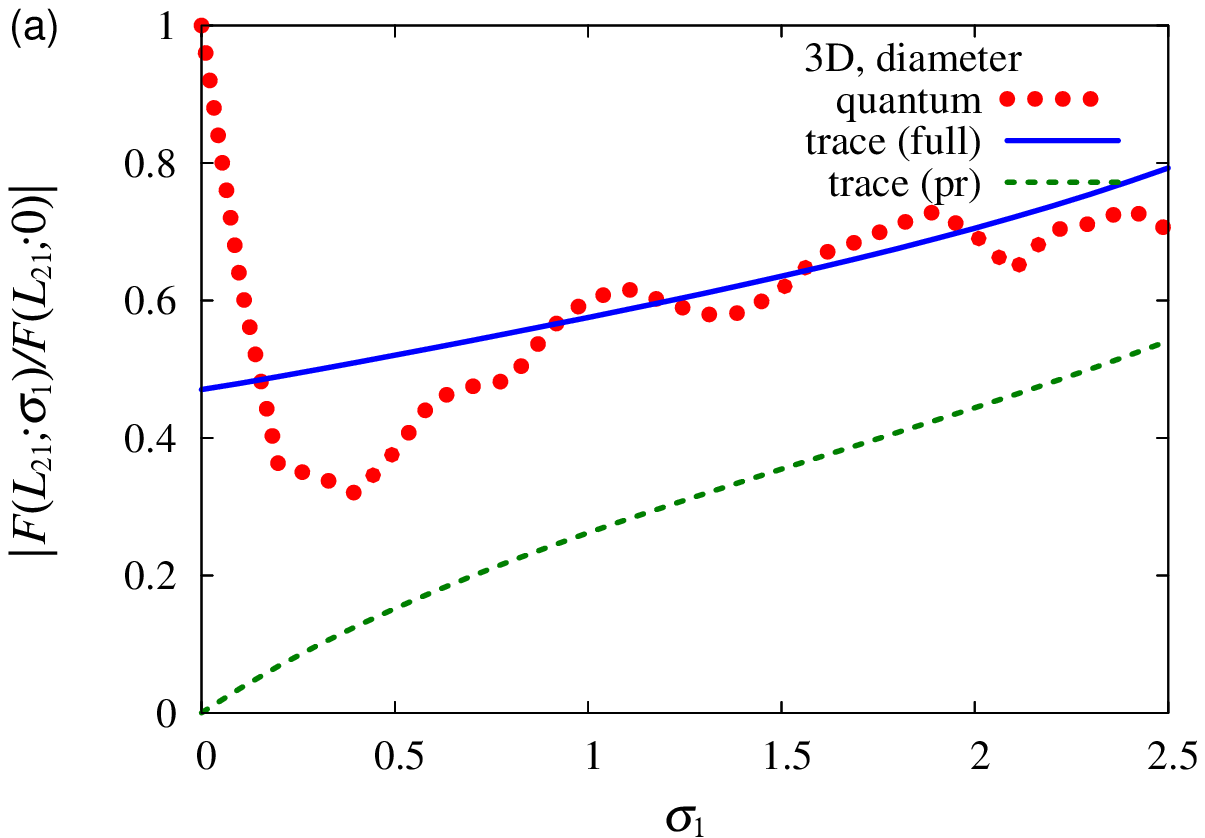}\\
\includegraphics[width=\linewidth]{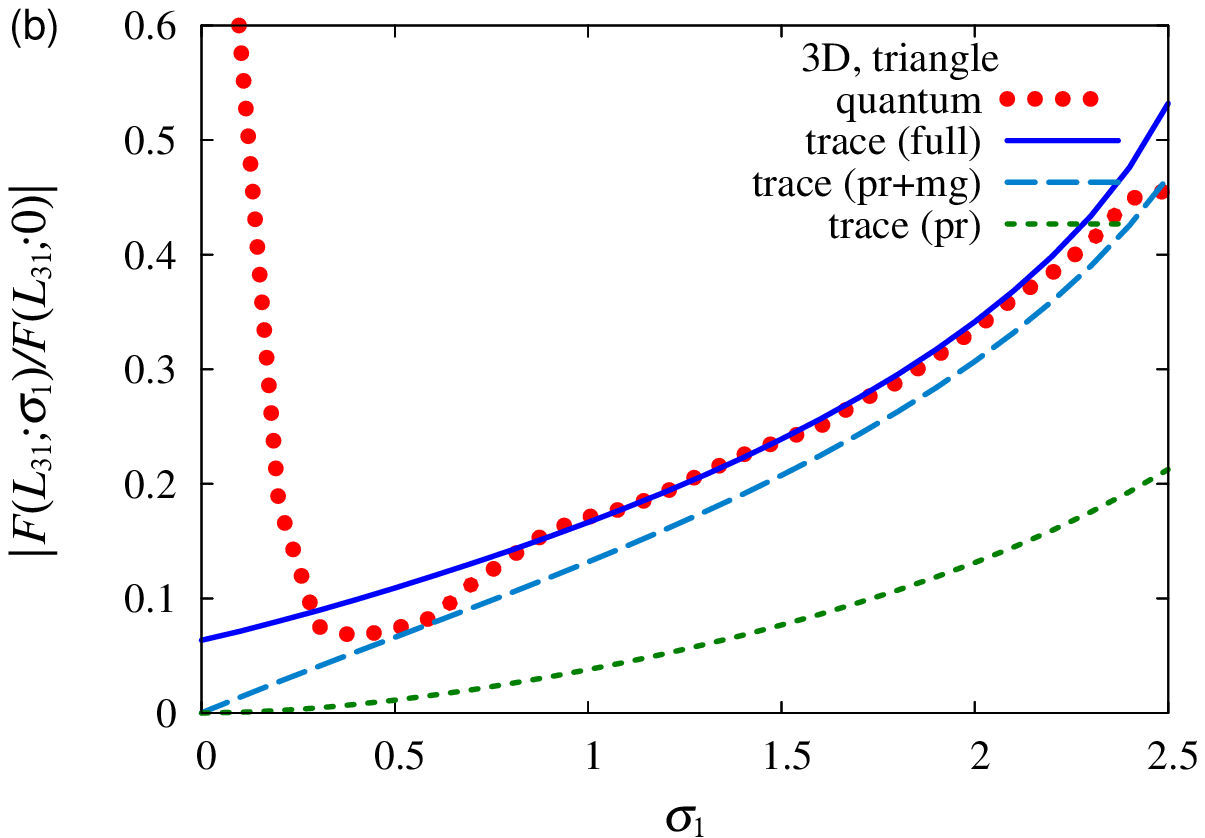}
\caption{\label{fig:ftl3d-obt} 
Same as Fig.~\ref{fig:ftlobt2d} but for the diameter (a) and triangle
(b) orbits in the three-dimensional TQS cavity potential.  In both
panels, the thick dotted (red) line represents the quantum result, the
solid (blue) line represents the result of the semiclassical trace
formula, taking account of all the marginal orbit contributions, and
the short-dashed line represents the contribution of the principal
family.  In the panel (b), the long-dashed (light blue) line
represents the result considering the principal family and the
marginal orbits, but without the secondary marginal orbits.}
\end{figure}

Quantum and semiclassical results for the Fourier amplitude
$|F(L_{pt})|$ at the lengths of classical periodic orbits $(p,t)$,
relative to those for the spherical cavity are shown in
Fig.~\ref{fig:ftl3d-obt}.
Similarly to the 2D billiard case, the principal term considerably
underestimates the quantum Fourier amplitude.  By taking into account
the marginal terms, quantum results are nicely reproduced for both
diameter and triangle orbits.  In the figure for diameter orbits,
ragged behavior of the quantum Fourier amplitude would be due to the
interference with the other periodic orbits, since the lengths of some
equatorial orbits on the neck surface cross with that of the the
prefragment diameter around $\sigma_1\sim 2$ as seen in
Fig.~\ref{fig:fmap}(b).

As shown in Fig.~\ref{fig:ftl3d-obt}(b), contribution of the
secondary-marginal orbit is considerably smaller than the principal
and marginal contributions.  The contributions of other one-parameter
families such as those shown in Fig.~\ref{fig:orbits} are expected to
be of the same order as the secondary-marginal family, and it may not
be so bad just to ignore them for simplicity.  It is also justified
from the Fourier spectrum shown in Fig.~\ref{fig:fmap}(a) where one
finds no significant peaks along those orbits.  Thus, one can consider
the semiclassical level density simply with the prefragment
periodic-orbit families.

\begin{figure}
\centering
\includegraphics[width=\linewidth]{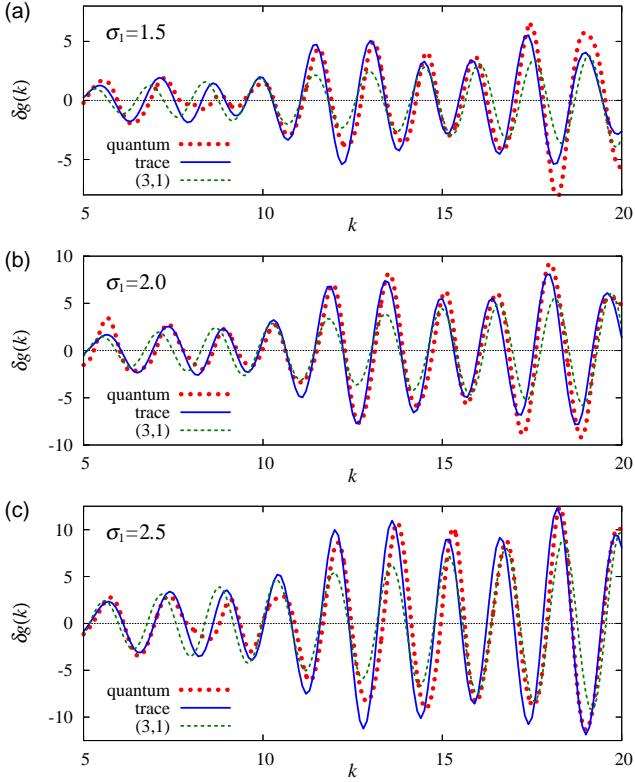}
\caption{\label{fig:sld3d} 
Oscillating part of the level density with averaging width $\gamma=0.3$
for several values of $\sigma_1$.  The semiclassical trace
formula taking account of three shortest prefragment orbit families
[solid (blue) line] is compared with the quantum level density [thick
dotted (red) line].  The contribution of the triangle orbit families
(3,1) [broken (green) line] is also shown.}
\end{figure}

In Fig.~\ref{fig:sld3d}, quantum level density is compared with the
semiclassical trace formula including the contributions of prefragment
diameter (2,1) and polygon families $(p,1)$ with $3\leq p\leq 5$.  In
these calculations, the averaging width is taken as $\gamma=0.3$.  For
every value of deformation $\sigma_1$, quantum results are nicely
reproduced by the contributions of those prefragment-orbit families.
One also sees that the quantum fluctuations are mostly attributed to
the contribution of the triangle family for large $\sigma_1$ where the
neck is well developed.  This can be understood from the fact that the
triangle orbits are the shortest among the families with the largest
degeneracy, and they also have the largest value of the occupation
rate $f_p$ among them, as shown in Table~\ref{table1}.

\section{Summary and conclusion}
\label{sec:summary}

Based on the Balian-Bloch formula, I have derived the contribution of
degenerate families of orbits confined in 2D truncated circular
billiard and 3D truncated spherical cavity systems.  In addition to
the truncated portion of the original families, contributions of the
marginal orbits should be considered independently as the end-point
correction to the former.  In applications to the 2D billiard and 3D
cavity potentials with TQS shape parametrization, those formulas have
been shown to successfully reproduce the quantum mechanical results.
Although the contributions of the marginal orbits are expected to play
minor roles in the semiclassical limit due to the lower degeneracies,
it turns out that they play a significant role in the shell effects in
the energy region of nuclei.  Their effect is important especially for
relatively small elongation where only a small portion of the
parameter space is occupied by the fully degenerate periodic-orbit
families, and it would be responsible for a prefragment shell effect
emerging at a rather early stage of the fission deformation process.

Using the semiclassical trace formula, shell effect associated with
the prefragments is extracted in a simple and natural way and can be
evaluated quantitatively through the nuclear fission processes.  The
periodic orbit theory with the formula derived in this work would thus
provide us a powerful tool to investigate the nuclear fission
dynamics.  Detailed analysis of the potential energy surface, taking
into account the asymmetric shape degree of freedom in the TQS cavity
model and discussions on the origin of asymmetric fission will be
presented in a separate paper\cite{AIM2018}.

\acknowledgments
The author thanks Dr.~Takatoshi Ichikawa and Prof.~Kenichi Matsuyanagi
for many fruitful discussions and valuable comments.  Some of the
numerical calculations were carried out on the computer system of
Yukawa Institute of Theoretical Physics.

\appendix
\renewcommand{\thesubsection}{\Alph{section}\arabic{subsection}}

\section{Trace formula for marginal orbits}
\label{sec:appA}

\subsection{Three-parameter family contribution}

For a regular polygon orbit $(p,t)$ in the three-dimensional spherical
cavity potential, the generic formula (\ref{eq:trace_gen}) can be
written as \cite{BB3}
\begin{gather}
g_{pt}(E)=\frac{2M}{\hbar^2}\Re\frac{\sin\varphi_{pt}}{\pi kR^{p-1}}
\left(\frac{ik}{4\pi}\right)^p\int dS_1\cdots dS_p e^{ikl_p}.
\label{eq:trace_trsph}
\end{gather}
In evaluating the surface integrals, the local surface coordinates
$(x_a,y_a)$ around the vertex $\mathrm{P}_a$ are defined as explained
in Sec.~\ref{sec:3D} (see Fig.~\ref{fig:local_coord}).

For the three-parameter family of polygon orbits, the integrals over
$x_1, y_1$, and $y_2$ should be exactly done, which gives the factor
\begin{equation}
\int dx_1dy_1dy_2=8\pi^2 R^3\sin(2\varphi_{pt}) f_p
\label{eq:int1}
\end{equation}
with the occupation rate $f_p$ given by Eq.~(\ref{eq:fp}).  The other
integrals are carried out using the SPA in the following way.  The
length of the orbit is expanded in $(2p-3)$-dimensional surface
coordinates $R\bm{u}=(x_2; x_3,y_3; \cdots; x_p,y_p)$ up to the
quadratic order as
\begin{gather}
l(1,\cdots,p)=L_{pt}+\frac{R}{4}\sum_{ab}K_{ab}u_au_b, \NN
K_{ab}=\frac{2}{R}\pp{^2 l_{pt}(u)}{u_a\partial u_b}.
\label{eq:l_expansion}
\end{gather}
$L_{pt}$ is the length of the orbit $(p,t)$.  $K$ represents the
curvature matrix for the orbit, which is dimensionless, symmetric, and
independent of the radius of the fragment.  Using the SPA, the
integrals are evaluated as
\begin{align}
&\int dx_2 dx_3dy_3\cdots dx_pdy_pe^{ikl(1\cdots p)} \NN
&=R^{2p-3}e^{ikL_{pt}}\int d^{2p-3}u\;
 \exp\left[\frac{ikR}{4}\sum_{ab}K_{ab}u_au_b\right] \NN
&=\frac{(4\pi iR/k)^{p-3/2}}{\sqrt{|\det K|}}e^{ikL_{pt}-i\pi n_-/2},
\label{eq:int2}
\end{align}
where $n_-$ denotes the number of negative eigenvalues of $K$.
Inserting (\ref{eq:int1}) and (\ref{eq:int2}) into
Eq.~(\ref{eq:trace_trsph}), one has
\begin{align}
& g_{pt}^{\rm(pr)}(E)
=\frac{2MR^2}{\hbar^2}\frac{f_p\sin\varphi_{pt}\sin(2\varphi_{pt})
 \sqrt{kR}}{\sqrt{\pi |\det K|}} \NN
&\quad \times \sin\left(kL_{pt}-\tfrac{\pi}{2}\mu_{pt}\right), \quad
\mu_{pt}=2p+n_-+\tfrac12.
\end{align}
The following relations have been checked numerically for each
$(p,t)$:
\begin{gather}
|\det K|=p\sin\varphi_{pt}, \quad
2p+n_-+\tfrac12=2t-p-\tfrac32 ~ (\mathrm{mod}~4)
\end{gather}
and $g_{pt}^{\rm(pr)}$ coincides with the Balian-Bloch formula
$g_{pt}^{\rm(sph)}$ for $f_p=1$ [see Eq.~(\ref{eq:BB_sph})].  Thus one
has the contribution of the three-parameter family as
\begin{align}
g_{pt}^{\rm(pr)}(E)&=f_p g_{pt}^{\rm(sph)}(E) \NN
&=\frac{2MR^2}{\hbar^2}f_p\sqrt{kR}A_{pt}^{\rm(sph)}
\sin(kL_{pt}-\tfrac{\pi}{2}\mu_{pt}), \\
A_{pt}^{\rm(sph)}
&=\frac{\sin\varphi_{pt}\sin 2\varphi_{pt}}{\sqrt{\pi|\det K|}}
=\sin2\varphi_{pt}\sqrt{\frac{\sin\varphi_{pt}}{\pi p}} \\
\mu_{pt}&=2t-p-\tfrac32,
\end{align}
where $A_{pt}$ represents the dimensionless amplitude which is
independent of $k$ and $R$.

\subsection{Contribution of the marginal orbit families with one
vertex on the joint}

To evaluate the contribution of the marginal family to the integral
(\ref{eq:trace_trsph}), any vertex can be put on the joint of the
spherical surface and the neighboring surface, and each gives an
identical contribution.  Thus, one can put the first vertex on the
joint and multiply it with $p$.  The marginal orbits form a
two-parameter family for polygon $(p>2t)$ and a one-parameter family
for diameter $(p=2t)$.  For a marginal polygon, the surface
coordinates $\xi_1$ and $y_2$ should be exactly integrated.
$\xi_1=R\varphi_1\sin\vartheta_A$ is related to the rotation about the
symmetry axis ($0\leq\varphi_1\leq 2\pi$), and
$y_2=R\psi\sin(2\varphi_{pt})$ to the rotation about the axis
$\mathrm{OP_1}$ with angle $\psi$ over the range $4\psi_p$.  The
integrations over these variables give the factor
\begin{equation}
\int d\xi_1 dy_2
=8\pi R^2\psi_p(\vartheta_A)\sin\vartheta_A\sin(2\varphi_{pt}).
\end{equation}
Integrating over the rest of $2p-2$ variables
$R\bm{u}=(\eta_1;x_2;x_3,y_3;\cdots;x_p,y_p)$ using the SPA by
expanding the length $l_p$ as (\ref{eq:l_expansion}), one has
\begin{align}
&\int dS_1\cdots dS_p e^{ikl(1\cdots p)} \NN
&=p\cdot\frac12\cdot 8\pi R^{2p}\psi_p(\vartheta_A)\sin\vartheta_A
 \sin(2\varphi_{pt}) \NN
&\quad\times e^{ikL_{pt}}\int d^{2p-2}u\,
 \exp\left[\frac{ikR}{4}\sum_{ab}K_{ab}'u_au_b\right] \NN
&=\frac{4\pi pR^{2p}\psi_p(\vartheta_A)\sin\vartheta_A\sin(2\varphi_{pt})
(4\pi i)^{p-1}}{(kR)^{p-1}\sqrt{|\det K'|}} \NN
&\quad\times e^{ikL_{pt}-i\frac{\pi}{2}n_-'}.
\end{align}
In the middle expression, the factor $p$ appears because any of the
$p$ vertices can be put on the joint as stated above.  The next factor
$1/2$ is to compensate for the integration range of $\eta_1$, which is
actually $\eta_1>0$ but extended to $(-\infty<\eta_1<\infty)$ by
assuming that the surface around the vertex $\mathrm{P_1}$ is the
extension of the neighboring wall outside the spherical surface.
$n_-'$ counts the number of negative eigenvalues of the
$(2p-2)$-dimensional curvature matrix $K'$.  In consequence, one
obtains
\begin{gather}
g_{pt}^{\rm(mg)}(E)=\frac{2MR^2}{\hbar^2}A_{pt}^{\rm(mg)}
 \sin\Bigl(kL_{pt}-\tfrac{\pi}{2}\mu_{pt}^{\rm(mg)}\Bigr),
\end{gather}
with
\begin{gather}
A_{pt}^{\rm(mg)}=\frac{p\psi_p(\vartheta_A)\sin\varphi_{pt}
 \sin(2\varphi_{pt})\sin\vartheta_A}{\pi\sqrt{|\det K'|}} \\
\mu_{pt}^{\rm(mg)}=2p+n_-'.
\end{gather}
The above equations are valid for a primitive polygon family.  For a
repeated polygon $m(p,t)$, the $j(p+1)$th vertex $(1\leq j < m)$ on
the joint can be placed either on the spherical surface or on the
neighboring surface.  Therefore, one should sum over all $2^{m-1}$
combinations (labeled $\beta$) for the choice of the surfaces:
\begin{gather}
\begin{split}
g_{m(p,t)}^{\rm(mg)}(E)&=\frac{2MR^2}{\hbar^2}
 \sum_\beta A_{m(p,t),\beta}^{\rm(mg)} \\
&\qquad \times\sin\Bigl(kmL_{pt}-\tfrac{\pi}{2}
 \mu_{m(p,t),\beta}^{\rm(mg)}\Bigr),
\end{split} \label{eq:mg_rep}
\end{gather}
with
\begin{gather}
A_{m(p,t),\beta}^{\rm(mg)}=
\frac{p\psi_p\sin\varphi_{pt}\sin(2\varphi_{pt})
 \sin\vartheta_A}{2^{m-1}\pi\sqrt{|\det K_\beta'|}}, \\
\mu_{m(p,t),\beta}^{\rm(mg)}=2mp+n_\beta'
\end{gather}

The marginal diameter orbits $(2t,t)=t(2,1)$ form a one-parameter
family generated by the rotation about the symmetry axis.  Their
contribution to the level density is derived in the same way as above,
and one has
\begin{gather}
g_{t(2,1)}^{\rm(mg)}(E)=\frac{2MR^2}{\hbar^2}
 \sum_\beta \frac{A_{t(2,1),\beta}^{\rm(mg)}}{\sqrt{kR}}
 \sin\Bigl(ktL_2-\tfrac{\pi}{2}
 {\mu}_{t(2,1),\beta}^{\rm(mg)}\Bigr),
\end{gather}
with
\begin{gather}
A_{t(2,1),\beta}^{\rm(mg)}=
\frac{\sin\vartheta_A}{2^{t-1}\sqrt{\pi|\det K_\beta'|}}, \\
\mu_{t(2,1),\beta}^{\rm(mg)}=n_\beta'-\tfrac12.
\end{gather}

\subsection{Marginal polygon family with two vertices on the joint}

For a polygon orbit family, there is the possibility of two vertices
being placed on the joint.  This forms a one-parameter family
according to the rotation about the symmetry axis.  The integration
with respect to $x_1=R\varphi_1\sin\vartheta_A$ gives the factor $2\pi
R\sin\vartheta_A$, and other $2p-1$ integrals are carried out by using
the SPA.  The curvature $K''$ is calculated under the assumption that
the two vertices are on the neighboring surface.  Taking account of
the $2p$ possible ways of selecting the two vertices on the joint, and
the extensions of surface integration ranges for the two surface
coordinates from $(0,\infty)$ to $(-\infty,\infty)$, one has
\begin{align}
&\int dS_1\cdots dS_p e^{ikl(1\cdots p)} \NN
&=2p\cdot\frac14\cdot 2\pi R^{2p}\sin\vartheta_A
 \frac{(4\pi i/kR)^{p-1/2}}{\sqrt{|\det K''|}}
 e^{ikL_{pt}-i\pi n_-''/2}.
\end{align}
Thus, the contribution to the level density is given by
\begin{gather}
g_{pt}^{\rm(mm)}(E)=\frac{2MR^2}{\hbar^2}\frac{A_{pt}^{\rm(mm)}}{\sqrt{kR}}
 \sin\left(kL_{pt}-\tfrac{\pi}{2}\mu_{pt}^{\rm(mm)}\right), \\
A_{pt}^{\rm(mm)}=
 \frac{p\sin\varphi\sin\vartheta_A}{2\sqrt{\pi |\det K''|}}, \\
\mu_{pt}^{\rm(mm)}=2p+n_-'' -\tfrac12
\end{gather}
for a primitive orbit family.  In cases of repeated orbits, one has to
consider all the possible combinations of the surfaces for
intermediate reflections on the joint as in Eq.~(\ref{eq:mg_rep}).

\subsection{Total contribution of the fragment-orbit family}

Summarizing the above contributions, the total contribution of the
orbit family $(p,t)$ confined in the spherical fragment is given by
\begin{align}
g_{pt}^{\rm(frag)}(E)
&=\frac{2MR^2}{\hbar^2}\sum_{D=1}^3 (kR)^{\frac{D-2}{2}}A_{pt}^{(D)} \NN
&\quad\times\sin\left(kL_{pt}-\tfrac{\pi}{2}\mu_{pt}^{(D)}\right),
\end{align}
where the summation over principal and marginal terms is expressed as
the sum over the degeneracy $D$.  The amplitudes and Maslov indices
are given by
\begin{gather}
\left\{\begin{array}{l@{\quad}l}
A_{pt}^{(3)}=f_p A_{pt}^{\rm(sph)}, &
\mu_{pt}^{(3)}=\mu_{pt} \\
A_{pt}^{(2)}=A_{pt}^{\rm(mg)}, &
\mu_{pt}^{(2)}=\mu_{pt}^{\rm(mg)} \\
A_{pt}^{(1)}=A_{pt}^{\rm(mm)}, &
\mu_{pt}^{(1)}=\mu_{pt}^{\rm(mm)}
       \end{array}\right.
\end{gather}
for a polygon $(p>2t)$, and
\begin{gather}
\left\{\begin{array}{l@{\quad}l}
A_{2t,t}^{(3)}=0 & \\
A_{2t,t}^{(2)}=f_2A_{2t,t}^{\rm(sph)}, &
\mu_{2t,t}^{(2)}=\mu_{2t,t}^{\rm(sph)} \\
A_{2t,t}^{(1)}=A_{2t,t}^{\rm(mm)}, &
\mu_{2t,t}^{(1)}=\mu_{2t,t}^{\rm(mm)}
       \end{array}\right.
\end{gather}
for a diameter.  The level density in terms of the wave-number
variable $k$ is expressed as
\begin{align}
&g_{pt}^{\rm(frag)}(k)
=\frac{\hbar^2k}{M}g_{pt}^{\rm(frag)}(E) \NN
&\quad =2R\sum_{D=1}^3(kR)^{D/2}A_{pt}^{(D)}
\sin\left(kL_{pt}-\tfrac{\pi}{2}\mu_{pt}^{(D)}\right)
\label{eq:gk_frag} \\
&\quad =2R\Im\left[\left\{\sum_{D}(kR)^{D/2}A_{pt}^{(D)}
e^{-i\pi\mu_{pt}^{(D)}/2}\right\}e^{ikL_{pt}}\right] \NN
&\quad \equiv 2R\Im\left[\mathcal{A}_{pt}(kR)e^{ikL_{pt}}\right].
\end{align}

\end{document}